\documentclass[superscriptaddress,secnumarabic,
amssymb,amsmath,nobibnotes,aps,prd,showkeys,showpacs,nofootinbib]{revtex4}%
\usepackage{graphicx}
\usepackage{epsf}
\usepackage{bm}
\usepackage{amsmath}
\usepackage{amsfonts}
\usepackage{amssymb}
\usepackage{epstopdf}
\usepackage{natbib}
\usepackage{color}
\usepackage{float}
\providecommand{\U}[1]{\protect\rule{.1in}{.1in}}

\newcommand{\be}{\begin{equation}}
\newcommand{\ee}{\end{equation}}

\newcommand{\mincir}{\raise
-3.truept\hbox{\rlap{\hbox{$\sim$}}\raise4.truept\hbox{$<$}\ }}
\newcommand{\magcir}{\raise
-3.truept\hbox{\rlap{\hbox{$\sim$}}\raise4.truept\hbox{$>$}\ }}

\newtheorem{remark}{Remark}[section]

\begin{document}


\title{Some topics in Cosmology \\
-- Clearly explained by means of simple examples --}


\author{Jaume  de Haro}
\email{jaime.haro@upc.edu}
\affiliation{Departament de Matem\`atiques, Universitat Polit\`ecnica de Catalunya, Diagonal 647, 08028 Barcelona, Spain}

\author{Emilio Elizalde}
\email{elizalde@ice.csic.es}
\affiliation{Institute of Space Sciences, Campus UAB, Carrer de Can Magrans s/n, 08193 Bellaterra (Barcelona), Spain}
\affiliation{International Laboratory for Theoretical Cosmology, TUSUR University, 634050 Tomsk, Russia}



\begin{abstract}
This is a very comprehensible review of some key issues in Modern Cosmology. Simple mathematical examples and analogies are used, whenever available. The starting point is the well know Big Bang Cosmology (BBC). We deal with the mathematical singularities appearing in this theory and discuss some ways to remove them. Next, and before introducing the inflationary paradigm by means of clear examples, we review the horizon and flatness problems of the old BBC model. We then consider the current cosmic acceleration and, as a procedure to deal with both periods of cosmic acceleration in a unified way, we study quintessential inflation. Finally, the reheating stage of the universe via gravitational particle production, which took place after inflation ended, is discussed in clear mathematical terms, by involving the so-called $\alpha$-attractors in the context of quintessential inflation.
\end{abstract}

\vspace{0.5cm}

 \par
\maketitle
\section{Introduction} 
A long time ago humans raised their eyes to the sky and started to try to understand everything that was around: the whole Universe. Its origin and evolution are among the greatest mysteries in the history of humanity. This quest was the beginning of a new science named Cosmology (from Cosmos, the Greek word for Universe). One of its main purposes is to  study the whole Universe chronology, which is related with very primordial questions for human beings: “where do we come from? where are we going?”. 

The first models to describe our Universe were proposed long time ago. Among them, the Cosmic Egg allegory that appeared in India in the 15th to 12th century B.C., which depicted a cyclic  Universe expanding and collapsing infinitely many times. Some centuries later the Greek philosophers  questioned themselves about the fundamental constituents of everything (the four elements), and then started to build some primitive models of the cosmos (Anaximander, Anaxagoras, Democritus, Aristotle, Aristarchus), up to the more elaborate Ptolemaic geocentric Universe. It was not until the 14th century A.C. that the Polish astronomer Nicolaus Copernicus, based in Aristarchus model --who had already got the idea right, many centuries ago, but had not been able to convince his contemporaries, nor future generations--  proposed an heliocentric Universe, later refined by Johannes Kepler. A milestone was laid when, in 1687, Isaac Newton published his influential {\it Principia}, where the Universe was described as static, steady state and infinite.

\
 
 But it was not until the second decade of the 20th Century that  Modern Cosmology appeared. This happened thanks to the new tools that allowed astronomers to obtain, for the first time in history, the positions and velocities of the celestial bodies (and thus treat the Universe as an ordinary physical system \cite{eliz21}), and thanks also to the solid theoretical basis provided by Albert Einstein's General Theory of Relativity (GR). His first model, proposed in 1917, was the well-known Einstein's static model \cite{Einstein} where,  for consistency, he introduced his now very famous cosmological constant (CC). The same year, Willem de Sitter \cite{Sitter} proposed another static and closed model describing a universe that was clearly expanding. But it was devoid of any matter or energy, only containing a CC. In spite of being a solution of Einstein's equations, it was generally considered as physically unrealistic.
 
 Five years afterwards, in 1922 and 1924,  the Russian mathematician Alexander Friedmann obtained families of solutions of the GR equations depicting expanding and contracting universes \cite{friedm12}. Later, in 1927,  the Belgian scholar Georges Lema{\^\i}tre, using the observational values for speeds and distances that the reputed astronomers  Vesto Slipher and Edwin Hubble had graciously provided him, was the first to propose that the Universe was actually expanding (however, in this model it had no origin in time, which extended from minus to plus infinity)  \cite{lemai27}. Anyway, the value he got for the expansion rate was quite close to the one Hubble would obtain two years later (namely, the famous Hubble constant). 
 
 \
 
 Once  the evolution of the Universe was clear, the question about its origin acquired relevant importance. That the Universe had had an origin was proclaimed by Lemaître \cite{lemai31}, a few years after his great discovery of the universe expansion. He even dared to propose a specific model for the beginning of the cosmos, as a tremendous explosion of a primeval atom ---an idea soon to be scientifically discredited, but that, owing to its extreme simplicity and its beautiful name, Big Bang, has persisted until now in the popular literature. Later, it was proven that, under very general conditions the GR equations for the Universe imply that it must have started from a "mathematical" singularity (since all GR solutions diverge at some finite value of time in the past), what is now termed the Big Bang (BB) singularity and is at the heart of the so-called Big Bang Cosmology (BBC). It is important to clarify from the very beginning that this singularity --although rigorous and unavoidable in principle-- is only mathematical (as most, if not all, singularities appearing in Physics). As usually happens, it just comes from the simplicity of the theory, and appears in a region where the model is no longer valid; e.g., when one assumes that GR holds at {\it all} scales, what is clearly not true: for it is not valid at small length scales. 
 
 Being more specific, when our patch of Universe (the one visible for us) was the size of an atom or less, as yet unknown physical effects, such as quantum effects (produced by a quantum field \cite{Fischetti} or by holonomy corrections as in LQC \cite{Ashtekar}) or non-linear curvature effects in the Hilbert-Einstein action \cite{Faraoni}, did play an important role and could eventually prevent the singularity from occurring. 
 
\

Moreover, the Big Bang Cosmology model had some shortcomings, notably the horizon and flatness problems, which could finally be overcome thanks to Alan Guth's brilliant idea of {\it cosmic inflation} \cite{Guth}.  This is only an implementation of the old BBC model, where at very early times, presumably at GUT scales, a very short period of extremely fast accelerated expansion of the Universe is assumed to have ocurred. In fact, 
 the inflationary paradigm, devised by Guth at the beginning of the 80's and improved by Andrei Linde and other cosmologists \cite{linde, {albrecht}},  is still nowadays considered as the simplest viable theory that describes the early universe in agreement with the most recent observations. It also has a  good   predictive power, because it is  able to explain the  origin of the present inhomogeneities in the Universe as quantum fluctuations during that epoch \cite{chibisov, starobinsky, pi, bardeen, Linde:1982uu}, matching well the  latest observational data from the Planck survey \cite{planck}.

\

A further leap forwar occurred when, towards the end of the last century and after very important hints had been accumulated in that direction (see, e.g., \cite{eliz21} for a detailed explanation), it was realized that the Universe expansion is actually accelerating 
\cite{riess, perlmutter}. Scientists try to explain this acceleration by introducing a new energy component: {\it dark energy}
 \cite{Copeland:2006wr}. Its most simple  realization is the re-introduction of Einstein's cosmological constant, what leads to the so-called $\Lambda$ Cold Dark Matter model (the nowadays standard cosmological model) \cite{eliz21}. However to match with the current observational data, the value of the CC in this model has to be fine tuned to an enormous precision. This is the reason why some  cosmologist introduced the idea of {\it Quintessence}, where an scalar field is responsible for the late-time acceleration,  as a new class of  dark energy (see \cite{tsujikawa}, for a review of these models).

\

In this way, one can think of Inflation  and Quintessence as of the two different sides of a coin or, after the introduction by 
 Peebles and Vilenkin, in their seminal paper \cite{pv}, of the concept of {\it Quintessential Inflation} (QI),   
as just a unique kind of  force, which  describes both inflation and the accelerated expansion.
The  idea of a unified picture of the Universe connecting the early and the present accelerated stages is possible,
 through the introduction of a single scalar field, also named the {\it inflaton}, as in standard inflation,  which at early times produces inflation while at late times provides quintessence. 
 
In addition, the majority of models in QI are so simple that they only depend on two parameters, which are determined by observational data.  And, since  the dynamical behavior during inflation is an attractor,  the dynamics of the model are obtained with the value of the scalar field and its derivative --initial conditions-- at some moment during this period; what shows the simplicity of quintessential inflation.

\

The assumption that the Universe had an inflationary phase readily implies that a reheating mechanism,  to match inflation with the Hot Big Bang universe, is needed \cite{Guth}. This is because the particles existing before the beginning of this period are extremely diluted by the end of inflation, resulting in a very cold universe. The most accepted idea to reheat the universe in the context of QI is through a phase transition, namely from inflation to kination (a phase where almost all the energy density of inflation turns into kinetic form \cite{Joyce}), the adiabatic regime being broken, which allows for an abundant production of particles. This mechanism is not unique, since a number of other alternatives can be used. Two of them seem to be the most efficient. The first  is  {\it Gravitational Particle Production} of massless particles, already studied long time ago in \cite{Parker,fmm,glm,gmm,ford,Zeldovich}, and  recently applied to quintessential inflation in \cite{Spokoiny, pv}, for massless particles, and also the  reheating via gravitational production of heavy massive particles conformally coupled to gravity \cite{kolb, kolb1,Birrell1, hashiba, J}. The second  mechanism is called {\it Instant Preheating} and was introduced in \cite{fkl}, and applied for the first time to  inflation in \cite{Felder}.  

\

The present review is organized as follows.
In Section II, we discuss the Big Bang model, and obtain first the fundamental equations in an easy way, by using a simplified version of the  Einstein-Hilbert action, together with the first law of thermodynamics. Once we have these equations, we deal with the Big Bang singularity and with future singularities, and we review some possible ways to remove them, such as in Loop Quantum Cosmology and in semi-classical Gravity.
Section III is devoted to the study of the inflationary paradigm. First of all, the famous horizon and flatness problems are described, to then introduce Alan Guth's proposal of cosmic inflation, which solves the shortcomings of  Big Bang Cosmology. We describe with great detail the slow-roll regime,  explaining in a clear way the conditions to get it, and its attractor nature. 
In Section IV, we consider the current cosmic acceleration, starting with Einstein's static  model of 1917, where he  introduced his celebrated cosmological constant.  This constant is now at the heart of the famous $\Lambda$ Cold Dark Matter model, which is presently the standard cosmological model, used by many scientists to depict our Universe. The importance of dark matter on a cosmological level, e.g. for the formation of large-scale structures  in  the Universe, cannot be underestimated. It can actually provide a better picture of the current state of the art of the Big Bang cosmological model (for easy to follow references, see \cite{dm1}).

 However, in such model the cosmological constant has to be very fine tuned, which is conceptually a  serious problem. This is the reason why other forms of {\it dark energy} have been introduced, in order to deal with this issue. 
One of these proposals is {\it quintessence}, which we review in the context of Quintessential Inflation: a theory that aims at unifying the early and the late time accelerated phases of our Universe. 
Reheating of the Universe is considered in Section V, where by using the quantum harmonic oscillator model, we introduce the well-know diagonalization method, with the aim to calculate the energy density of particles created  via  gravitational particle production. As an application, we calculate the reheating temperature in the so-called $\alpha$-attractor models, which derive rather naturally from fundamental super-gravity theories. Finally, in the last Section some relevant historical notes are included, which describe a few crucial moments and important developments in the history of Modern Cosmology. A much more detailed account of these historical issues is provided in the recent book by one of the authors \cite{eliz21}, which is a perfect complement to the present review. Here, a much more specific, technically solid and detailed quantitative explanation of the concepts will be given, with all the relevant formulas and with the help of many comprehensible examples.


\section*{Conventions}

Throughout the work, natural units are used, $\hbar=c=k_b=1$, where $\hbar$ is the reduced Planck constant, $c$ the velocity of light, and $k_b$  Boltzmann's constant.  In {\it natural} units, one has:
$$[energy]=[mass]=[temperature]=[length]^{-1}=[time]^{-1}.$$

\

In these units, the reduced Planck mass reads $M_{pl}=\sqrt{\frac{1}{8\pi G}}\cong 2.44\times 10^{18}$ GeV, being $G$ { Newton's} constant.

 \
 
 Planck's scale:
 
 \begin{enumerate}
\item { Planck's} lenght:  $l_{pl}=\left(  \frac{G\hbar}{c^3}\right)^{1/2}=1.616\times 10^{-33}$ cm. { Compare with Bohr's radius: $r_B\cong 5.3\times 10^{-9}$ cm.}

\item { Planck's} time:  $t_{pl}=\frac{l_{pl}}{c}=5.391\times 10^{-44}$ s.

\item { Planck's} mass:  $m_{pl}=\left(  \frac{\hbar c}{G}\right)^{1/2}=2.117\times 10^{-5}$ g. {  Compare with the proton mass : $m\cong 1.7\times 10^{-24}$ g. }

\item { Planck's} temperature:  $T_{pl}=\frac{m_{pl}c^2}{k_B}\cong 1.4\times 10^{32}$ K.  { The current temperature of our Universe is $2.73$ K. The temperature of the solar surface is  approximately $6,000$ K.}

\item { Planck's} energy density $\rho_{pl}=\frac{m_{pl} c^2}{l^3_{pl}}= 4.643\times 10^{114}$ erg. $\mbox{cm}^{-3}$.

\

The energy, mass and temperature are either given in GeV or in terms of  $M_{pl}$. For example,  the temperature of the Universe one second after the Big Bang is around $10^{-3}$ GeV  or $10^{-21}  M_{pl}$, which  in IS units is approximately   $10^8$ K.

\end{enumerate}

\section{Big Bang Cosmology}

Modern Cosmology is based on Einstein's Equations (EE) of General Relativity (GR), which relate the geometry of spacetime -a four dimensional manifold- with the matter/energy it contains, in the following way: {\it "matter tells spacetime how to curve and spacetime tells test particles how to move"}. A test particle moves in   spacetime along a geodesic curve.

\

Being more precise, 
given a distribution of masses, the unknown variables are the coefficients of the metric $g_{\mu\nu}$,  and 
the equations of {GR} are second-order  partial differential equations (PDE)  containing some constraints (which are equations of order one or zero). 
The most important of them is the Hamiltonian constraint:
the total Hamiltonian of the system (the Universe) vanishes. These equations are
{ \begin{eqnarray}
R_{\mu\nu}-\frac{1}{2}Rg_{\mu\nu}=\frac{8\pi G}{c^4}T_{\mu\nu}, \qquad \mu,\nu=0,1,2,3,
\end{eqnarray}}
where
$R_{\mu\nu}=R^{\alpha}_{\mu\alpha\nu}$ is the { Ricci} tensor, $R=R_{\mu}^{\mu}$ the { Ricci} scalar (or scalar curvature) and $T_{\mu\nu}$ is the stress-energy tensor (containing all the information about the masses and energies of the Universe).

\

Here 
\begin{eqnarray}
R^{\rho}_{\sigma\mu\nu}=\partial_{\mu}\Gamma^{\rho}_{\nu\sigma}-\partial_{\nu}\Gamma^{\rho}_{\mu\sigma}+\Gamma^{\rho}_{\mu\lambda}\Gamma^{\lambda}_{\nu\sigma}
-\Gamma^{\rho}_{\nu\lambda}\Gamma^{\lambda}_{\mu\sigma},
\end{eqnarray}
is the { Riemann} curvature tensor, where 
\begin{eqnarray}
\Gamma^{\rho}_{\mu\nu}=\frac{1}{2}g^{\mu\lambda}\left( \partial_{\nu}g_{\lambda\mu}+\partial_{\mu}g_{\lambda\nu}-\partial_{\lambda}g_{\mu\nu}\right),
\end{eqnarray}
are the {Christoffel} symbols.

 \
 
 Fortunately, in Cosmology these equations simplify very much. In fact, we just know about
our patch of the Universe, that is, the part of the Universe that is visible to us. 
({Recall that, owing to inflation and also to the finite velocity of light, the whole Universe is not visible to us}). The extension of our patch is of the order of $3,000$ Mpc ($1$ Megapasec (Mpc)
$=3.26\times 10^6$ light years $=3.08 \times 10^{19}$ Km), and it is observed to be very
homogeneous (we see the same properties at all places of our patch) and isotropic (it does not matter which direction our telescope is pointing to, properties also remain the same), on scales larger than say $100$ Mpc,
but it does exhibit inhomogeneous structures on smaller scales (as, e.g., on scales of the Milky Way).

\

So, working on large scales, we can safely assume an homogeneous an isotropic Universe, which enormously simplifies the corresponding EE: the only variable is now the cosmic time (spatial coordinates do not appear in the equations due to  homogeneity). Thus, we are left with a set of {ordinary differential equations} (ODE), instead of the original PDE. The hypothesis of homogeneity and isotropy are extremely well supported by surveys of galaxies (e.g. SDSS) and by the recent results of the WMAP and the Planck satellites
(for updated information the reader can look at the following references \cite{homo-iso1}).

 \subsection{Hubble's law}
 
 {One of the most important discoveries in human history is the fact  that} our Universe is {expanding}. This conclusion was reached  through the {cosmic {Doppler} effect}: light of distant celestial objects is red-shifted, i.e., owing to the Universe expansion, the wavelength of the light emitted by these objects grows (the waves decompress), so the light we see from far away galaxies is displaced towards the red region of the spectrum. On the contrary, in an hypothetically contracting Universe, the wave-length of the emitted light would get compressed, and we would see a displacement towards the blue. This is what actually happened with the very first measurement 
performed by Vesto Slipher, in 1912, corresponding to the nearby galaxy Andromeda. It was a blueshift, due to the fact that in this case gravitational attraction is much larger than the effect of the local expansion of the Universe. And, in fact, Andromeda is approaching the Milky Way at a very high speed: the two galaxies will collide in the future. This is an important lesson we need to learn: cosmic expansion always competes with the gravitational forces of highly massive objects around, which provide contributions to the total redshift that astronomers have to disentangle (what in general is quite difficult to do).
{ You can see a very nice explanation of the  Doppler effect, in the context of GR,  in episode 8 of 
Carl Sagan's "Cosmos"  series} \cite{CScosmos}.

\

One of the fundamental principles of modern cosmology is Hubble’s empirical law (1929) \cite{hubble29}, which he obtained by measuring (mainly
by using Henrietta Leavitt’s law for Cepheid variable stars) the distances to spiral nebulae (given in Mpc) and comparing them
with the table of speeds (measured in kms), obtained by Vesto Slipher a few years before, and by then already published
in Eddington’s aclaimed book \cite{eliz21}. Actually, Hubble’s law had been obtained by Georges Lema{\^\i}tre two years before, a fact that was recently recognized, at last. As a result, the expansion law is now officially called the Hubble-Lema{\^\i}tre law,
 which relates the relative velocities of observers, as follows:
 In an expanding, homogeneous and isotropic Universe, 
the velocity of the observer $B$ with respect to $A$  is
\begin{eqnarray}\label{Hubblelaw}
\vec{v}_{AB}=H(t)\vec{r}_{AB}  \Longleftrightarrow \dot{\vec{r}}_{AB}=H(t)\vec{r}_{AB},\end{eqnarray}
where $H(t)>0$ is the so-called {{ Hubble rate}} and $\vec{r}_{AB}$  is the vector pointing from $A$ towards $B$.

\

Integrating this equation, one gets 
$
\vec{r}_{AB}(t)=a(t)\vec{r}_{AB}(t_i),$
where the dimensionless function $a(t)=e^{\int_{t_i}^t H(s)ds}$ is the so-called {{\it  scale factor}}. What is important is to note that $d_{AB}(t_i)=|\vec{r}_{AB}(t_i)|$ is the distance, at time $t_i$,   between points $A$ and $B$, and that at time $t$ this distance is $d_{AB}(t)> d_{AB}(t_i)$  (for $t>t_i$, because $H(t)>0\Longrightarrow a(t)>1$), so the Universe is expanding. 
 
 \
 
  In fact,  the distance between two points at different epochs is given by the formula { $d_{AB}(t_2)= \frac{a(t_2)}{a(t_1)}d_{AB}(t_1)$}, that is, the scale factor tells us how the distance between two points scales with time. To get a clear idea of this expansion, we can "imagine", as the simplest model,  that the Universe is an {inflating balloon} --in more rigorous mathematical terms, the 3-dimensional spherical surface of a 4-dimensional ball, with time being the { radial} coordinate-- and  $r_i$ its radius at time $t_i$, then the radius of the balloon at time $t$ would be 
$a(t)r_i$.  
In addition, one has  $H(t)=\frac{\dot{a}(t)}{a(t)}$, which means that this quantity  is the expansion rate of the Universe, where reliable observational data tell us that its current value, $H_0$, is approximately 
$70$  $\frac{{\mbox{km/s}}}{{\mbox{Mpc}}}.$

\

Before closing this point, we have to say that $H_0$ is both the most important cosmological parameter and also the most difficult to calculate --both because of the difficulty in measuring cosmological distances reliably, and of disentangling the cosmological contribution from other components of the observed redshift of a celestial object. The calculated value of $H_0$  has been changing a lot, from the time of the first reported measurements by Lema{\^\i}tre (600) and Hubble (500), down later to a value of 50, the preferred one some decades ago \cite{evolH0}. Even presently, there is a new and quite sharp controversy among different groups (see, e.g., \cite{ekom20} and references therein), what presently sets this value between 67 and 74, according to the results reported by different groups (there is namely over a 10\% discrepancy, even right now!) \cite{eliz21}. This issue has been named the expansion rate tension. This is  a very important issue nowadays and we would like to give a bit more information regarding the nature of this tension, which mainly affects the late-time versus the early-time data results. To give a detailed account of this issue lies beyond the scope of the present paper, but a very well documented version of it can be found, e.g. in 
\cite{tension1}
\

 \subsection{The cosmic equations}

 The main ingredient to obtain the dynamical equations of the cosmos at large scale is the 
scalar curvature or { Ricci} scalar, which  for a spatially flat  space-time (later we will see that our Universe could be spatially closed, flat or open), is given by 
$R(t)=6(\dot{H}(t)+2H^2(t))$.  In addition, we  assume that, at large scales, galaxies can be taken as particles of an homogeneous  fluid filling the Universe, whose energy
density is 
denoted by $\rho(t)$. The corresponding Lagrangian, in natural units,  was given by {Hilbert} in 1915, immediately after { Einstein} had { postulated  "a la Newton" } the equations of { GR} (see  the historical note at the end) 
\begin{eqnarray}\label{HElagrangian}{\mathcal L}(t)=V(t)\left(\frac{R(t) M_{pl}^2}{2}-\rho(t)\right),\end{eqnarray}
where $V(t)\equiv a^3(t)$ is the {\it volume} and $M_{pl}$   is the reduced {Planck} mass.
 
 \
 
The corresponding {Hilbert-Einstein's} action reads $S(t)=\int_{t_{i}}^t {\mathcal L}(s) ds$. {Then, since 
\begin{eqnarray}{\mathcal L}=-\frac{\dot{V}^2 M_{pl}^2}{3V}+{\ddot{V}M_{pl}^2}-\rho V,
\end{eqnarray}
and  ${\ddot{V}M_{pl}^2}= 
\frac{d}{dt}\left( {\dot{V}M_{pl}^2}  \right)$ is a total derivative, it can be disregarded because it does not have any
influence in the dynamical equations. Thus, we will use the following Lagrangian, which only contains first order derivatives with respect to the volume variable:
\begin{eqnarray}\label{HElagrangian1}\bar{\mathcal L}=-\frac{\dot{V}^2 M_{pl}^2}{3V}-\rho V =-3H^2M_{pl}^2V- \rho V.
\end{eqnarray}

\

 \begin{remark}
{ At this point we should recall that in classical mechanics,
given a Lagrangian of the form ${\mathcal L}(x,\dot{x},t)$, performing the variation of the action $S(t)=\int_{t_i}^t {\mathcal L}(x(s), \dot{x}(s), s)ds$ with respect to the dynamical variable $x$, one gets the so-called {Euler-Lagrange} equation
\begin{eqnarray}
\frac{d}{dt}\left(\frac{\partial {\mathcal L}}{\partial \dot{x}} \right)= \frac{\partial {\mathcal L}}{\partial {x}},
\end{eqnarray}
whose equivalent formulation is to consider the Hamiltonian, obtained via the { Legendre} transformation
\begin{eqnarray}\label{hamiltonian}{\mathcal H}=\dot{x}p_x-{\mathcal L}, \qquad \mbox{where} \qquad p_x\equiv \frac{\partial {\mathcal L}}{\partial \dot{x}}.\end{eqnarray}
Then, the {Euler-Lagrange} equations are equivalent to the { Hamiltonian} equations:
\begin{eqnarray}
\dot{x}= \partial_{p_x}{\mathcal H},\qquad \dot{p}_x= -\partial_{x}{\mathcal H}.\end{eqnarray}

}
\end{remark}

 \

Coming back to our Lagrangian,
the { Legendre} transformation leads to :
\begin{eqnarray}\label{hamiltonian}{\mathcal H}=\dot{V}p_V-\bar{\mathcal L},\end{eqnarray}
where  the corresponding momentum is $p_V\equiv \frac{\partial \bar{\mathcal L}}{\partial \dot{V}}= -\frac{2\dot{V} M_{pl}^2}{3V}$. Then, a simple calculation shows that the Hamiltonian is given by
\begin{eqnarray}\label{hamiltonian1}{\mathcal H}= -\frac{\dot{V}^2M_{pl}2}{3V}+\rho V =-3H^2M_{pl}^2V+\rho V.   \end{eqnarray}

As we have already explained,  the EE contain some constraints, and one of them states that the Hamiltonian vanishes. So, we have
\begin{eqnarray}\label{friedmann} { H^2=\frac{\rho}{3M_{pl}^2}},\end{eqnarray}
which is the well-known  { Friedmann equation} ({FE}).

 \begin{remark}
 The Friedmann equation could also be obtained using the time $Nds=dt$, where $N$ is the so-called {\it lapse function}. From this new time, the Lagrangian becomes
 \begin{eqnarray}
 \bar{\mathcal L}=-3\frac{\tilde{H}^2}{N}M_{pl}^2V- \rho VN,
 \end{eqnarray}
 where $\tilde{H}=\frac{1}{a}\frac{da}{ds}=NH$.
 
 \
 
 Then, the variation with respect to the lapse function yields
 \begin{eqnarray}
 \frac{\partial {\mathcal L}}{\partial N}=0 \Longrightarrow
 H^2=\frac{\tilde{H}^2}{N^2}=\frac{\rho}{3M_{pl}^2}.
 \end{eqnarray}
 \end{remark}
 
 \
 
Once we have obtained the constraint,  we need to find the dynamic equation. To do that, we need {\it  the first law of thermodynamics}}. Assuming an adiabatic evolution of the Universe, i.e., that the 
Universe entropy is conserved, one has
\begin{eqnarray}\label{conservation0}{d(\rho V)=-PdV},
\end{eqnarray}
where $\rho$ is, once again, the energy density of the fluid  and  $P$  its pressure. Taking the derivative with respect to the cosmic time $t$, it can be expressed as a {conservation equation} ({CE})
\begin{eqnarray}\label{eqnarray} { \dot{\rho}=-3H(\rho+ P)}.\end{eqnarray}

Next, 
taking the derivative of the {FE} and using the {CE}, one obtains the so-called {Raychaudury Equation} ({RE})
\begin{eqnarray}\label{raychaudury} {\dot{H}=-\frac{1}{2M_{pl}^2}(\rho+P)}.\end{eqnarray}
Finally, to obtain the dynamics of the Universe we need the
relation between the pressure and the energy density, i.e., the
{ equation of state} ({EoS}), which for the moment we assume that it has the simple form $P=P(\rho)$ .

 \
 
 \begin{remark}
  { We should note that the {Raychauduri} equation can also be obtained from the {Euler-Lagrange} one. In fact, a simple calculation leads to 
\begin{eqnarray}
\frac{d}{dt}(\partial_{\dot V} \bar{\mathcal L} )=-2\dot{H}M_{pl}^2.
\end{eqnarray}
On the other hand, 
\begin{eqnarray}
\partial_{V}\bar{\mathcal L} =3H^2M_{pl}^2-\partial_V(\rho V),\end{eqnarray}
thus, using the {Friedmann} equation and the first law of thermodynamics, one gets 
\begin{eqnarray}
\partial_{V}\bar{\mathcal L} =\rho +P,\end{eqnarray}
what leads to the {Raychauduri} equation.}
 \end{remark}

 \
 
 Summing up, {{\it the constituent equations}} in cosmology are: 

$$H^2=\frac{\rho}{3M_{pl}^2}  \quad \mbox{ Friedmann Equation}.$$
$$ \dot{\rho}=-3H(\rho+ P) \quad \mbox{ Conservation Equation}.$$
$$\dot{H}=-\frac{1}{2M_{pl}^2}(\rho+P)  \quad \mbox{ Raychaudury Equation}. $$
$$P=P(\rho) \quad \mbox{ Equation of State}. $$

Note that the variables are  $H$ and {$\rho$}, but they are related via the {FE}. So, in practice we have only one unknown variable. Moreover, the {CE} and the {RE} are equivalent, what means that we only need to solve one of them. For example, the {CE}, which can be written as 
\begin{eqnarray}
{ \dot{\rho}=-\frac{\sqrt{3\rho}}{M_{pl}}\left(\rho+ P(\rho)\right)},
\end{eqnarray}
which is only a simple and solvable first order differential equation.

 \

Next,  we define the EoS parameter
as $w_{eff}=\frac{P}{\rho}$, which using the constituent equations, can be written as follows
\begin{eqnarray}\label{EoS}
\qquad w_{eff}=-1-\frac{2\dot{H}}{3H^2}.
\end{eqnarray}
Then, from the EoS parameter, the
  {\it acceleration equation} could be written as follows  \begin{eqnarray}\label{acceleration} 
\frac{\ddot{a}}{a}=\dot{H}+H^2=-\frac{{H}^2}{2}(1+3w_{eff}), \end{eqnarray}
and thus,  on can conclude that the Universe is {decelerating} for $w_{eff}>-1/3$ and that it is {accelerating} for $w_{eff}<-1/3$.
In addition, from the Raychaudury equation, for $w_{eff}>-1$ the Hubble parameter decreases, and for $w_{eff}<-1$ 
({phantom fluid}) it increases. 

\

Finally,  from the first law of thermodynamics one can easily show that for $w_{eff}=w=\mbox{constant}$, the energy density scales as $\rho\propto a^{-3(w+1)}$.

 \subsection{Singularities}

 We consider the following linear {EoS} $P=w\rho$, where $w>-1$ ({non phantom fluid}) is a constant {EoS} parameter ($w=0$ for a dust fluid and $w=1/3$ for radiation).  The combination of the {FE} and the {RE} leads to
  \begin{eqnarray}\dot{H}=-\frac{3}{2}(1+w)H^2,
  \end{eqnarray}
whose solution is given by
\begin{eqnarray}\label{BB}
H(t)=\frac{H_0}{\frac{3}{2}(1+w)H_0(t-t_0)+1}=\frac{2}{{3}(1+w)(t-t_s)},
\end{eqnarray}
 where $t_0$ is the present time, $H_0$ the current value of the Hubble rate, and  $t_s=t_0-\frac{2}{{3}(1+w)H_0}$ is the time when the singularity appears. Inserting this expression in the {FE}, one gets
 \begin{eqnarray}{\rho(t)=\frac{4 M_{pl}^2}{{3}(1+w)^2(t-t_s)^2}} .\end{eqnarray}

 \
 
 So, we see that the solutions, $H(t)$ and $\rho(t)$, diverge at time $t_s<t_0$. This  is the so-called {{Big Bang}} ({BB}) singularity (see the historical at the end to understand where that name comes from), where the scale factor  
\begin{eqnarray}a(t_s)=\lim_{t\rightarrow t_s}a(t_0)e^{-\int_{t}^{t_0}H(s)ds}=0,
\end{eqnarray}
that is,
the "radius" of the Universe vanishes at the {BB} singularity.

\

On the other hand, {the age of our Universe} is approximately given by
\begin{eqnarray} 
t_0-t_s=\frac{2}{{3}(1+w)H_0}\cong {14  \mbox{ {billion years}}},\end{eqnarray}
 where we have used that the current value of the Hubble rate is  $H_0\cong 70$  $\frac{{\mbox{km/s}}}{{\mbox{Mpc}}}$ and $w\sim 0$.

 \
 
 Here, it is important to remark that there are other kind of mathematical singularities. In fact, 
 when the {EoS} is non-linear, as 
 $P=-\rho+A\rho^{\alpha}$, 
 one can encounter these different future singularities \cite{Tsujikawa}:
\begin{enumerate}
\item Type I (Big Rip): For $t\rightarrow t_s$, $a\rightarrow \infty$, $\rho\rightarrow \infty$ and $|P|\rightarrow \infty$.
\item Type II (Sudden): For $t\rightarrow t_s$, $a\rightarrow a_s$, $\rho\rightarrow \rho_s$ and $|P|\rightarrow \infty$.
\item Type III (Big Freeze): For $t\rightarrow t_s$, $a\rightarrow a_s$, $\rho\rightarrow \infty$ and $|P|\rightarrow \infty$.
\item Type IV (Generalized sudden): For $t\rightarrow t_s$, $a\rightarrow a_s$, $\rho\rightarrow 0$, $|P|\rightarrow 0$, and
higher-order derivatives of $H$ diverge.
\end{enumerate}

 \subsubsection{Big Rip singularity}
This singularity is the future equivalent to the Big Bang one, and it is obtained, for instance, when one deals with a phantom fluid with a linear EoS ($w<-1$). The solution is given by
(\ref{BB}), but now $t_s>t_0$, that is, the singularity appears at late times.

\subsubsection{Sudden singularity}

In \cite{Barrow} Barrow proposed a new kind of finite time future singularity appearing in an expanding FLRW universe. The singularity may also show up without violating the strong energy condition: $\rho> 0$ and $3 P+ \rho >0$. This singularity was named  the {\it sudden singularity}.   

To deal with this kind of singularities,
we consider a nonlinear EoS $P=-\rho-f(\rho)$. In this case the   conservation equation becomes
$\dot{\rho}=3Hf(\rho) $, and using the Friedmann equation, one gets
\begin{equation}
    \dot{\rho}=\sqrt{3}\frac{\rho^{1/2}}{M_{pl}}f(\rho).
\end{equation}

Choosing, as in \cite{Nojiri}, $f(\rho)=\frac{A M_{pl}}{\sqrt{3}}\rho^{\nu+\frac{1}{2}}$, where
$A$ and $\alpha$ are two parameters, one obtains the first-order differential equation
\begin{equation}
    \dot{\rho}=A\rho^{\nu+1}, 
\end{equation}
whose solution is 
\begin{equation}
\rho(t)=\left\{\begin{array}{cc}
   (\rho_0^{-\nu}-\nu A(t-t_0))^{-1/\nu}  & \nu\not= 0 \\
  \rho_0e^{A(t-t_0)}   & \nu=0,
\end{array}  \right.  
\end{equation}
where $\rho_0$ is the current value of the energy density.

Now, we consider the case $\nu<-1/2$. The Hubble rate is given by
\begin{equation}\label{hubble}
H(t)= \frac{1}{\sqrt{3}M_{pl}}\left(\rho_0^{-\nu}-\nu A(t-t_0)\right)^{-1/2\nu},    
\end{equation}
which, introducing the time $t_s=t_0+\frac{\rho_0^{-\nu}}{ \nu A}$, can be written as 
\begin{equation}
H(t)= \frac{1}{\sqrt{3}M_{pl}}\left(\nu A(t_s-t)\right)^{-1/2\nu},   \end{equation}
and thus, the scale factor is given by
\begin{equation}\label{scalefactor}
    \ln\left(\frac{a(t)}{a_s} \right)=-\frac{2}{3M_{pl}A(2\nu-1)}
    \left(\nu A(t_s-t)\right)^{(2\nu-1)/2\nu}
\end{equation}

\

Then, for a non phantom fluid, namely for $A<0$ (the effective EoS parameter is $w_{eff}\equiv \frac{P}{\rho}=-1-\frac{A M_{pl}}{\sqrt{3}}\rho^{\nu-\frac{1}{2}}>-1$), we  see that
$t_s>t_0$, and since $\rho(t_s)$  vanishes, we find that the pressure $P$ diverges at the instant $t_s$, yielding a future sudden singularity.

\

\subsubsection{Big Freeze singularity}
For the same EoS as in the previous subsection,
here we consider the case $\nu>1/2$ and $A>0$, that is, a phantom fluid, what implies that $t_s>t_0$. From (\ref{scalefactor}) we see that $a(t)\rightarrow a_s$ when $t\rightarrow t_s$, and from (\ref{hubble}) we deduce that both the energy density and the pressure diverge at that instant, which leads to a Big Freeze singularity.

\

\subsubsection{Generalized Sudden singularity}

We continue with the same EoS, but with $-1/2<\nu<0$ and $A<0$ (non-phantom fluid). 

In this situation, $t_s>t_0$ and, once again, the scale factor converges to  $a_s$ when $t\rightarrow t_s$; but now the energy density and the pressure go to zero as the cosmic time approaches $t_s$. In addition, looking at the Hubble rate obtained in the formula (\ref{hubble}), we easily conclude that when $-\frac{1}{2\nu}$ is not a natural number then some high order derivative of the Hubble parameter will diverge at $t=t_s$, thus obtaining a Generalized Sudden singularity.

\

To end this section, note that the case $0<\nu<1/2$ and $A<0$ (non-phantom fluid) corresponds to a Big Bang singularity and, for $0<\nu<1/2$ and $A>0$ (phantom fluid), one gets a Big Rip singularity. In addition, when $\nu=0$ and $A>0$ one obtains the so-called Little Rip \cite{Frampton, Framptona}, where $w<-1$ and asymptotically converges to $-1$. In fact, in this case the energy density is given by
$\rho(t)=\rho_0e^{A(t-t_0)}$, which diverges when $t\rightarrow \infty$. So, 
\begin{equation}
    w=\frac{P}{\rho}=-1-\frac{AM_{pl}}{\sqrt{3\rho}}\rightarrow -1.
\end{equation}

Finally, it remains the case $\nu=1/2$ and $A>0$, where the Hubble rate is given by 
\begin{equation}
    H=\frac{2}{\sqrt{3}M_{pl}A (t_s-t)},
\end{equation}
and thus, the scale factor is given by
\begin{equation}
    \ln\left(\frac{a(t)}{a_0} \right)=-\frac{2}{\sqrt{3}M_{pl}A}\ln\left(\frac{t_s-t}{t_s-t_0}\right),
\end{equation}
what means that the scale factor diverges when $t\rightarrow t_s$, and thus, we obtain a Big Rip singularity.

  \

 \subsection{Removing singularities}
 It is very important  to realize  that
 the singular solutions we have found are just mathematical solutions of the EE. And we know that
 GR is a viable  theory that has been proven to match the observational data at low energy densities, but we still do not know what are the valid physical laws
 at very high scales. Recall that, at Planck scales, $\rho_{pl}\sim M_{pl}^4\sim 10^{114} \mbox{ erg}/ {\mbox{cm}}^{3}$ and $l_{pl}\sim M_{pl}^{-1}\sim  10^{-33}$ cm. 
 
 \
 
 Taking this into account, and the very clear fact that the EE --as it happens with Newton's ones-- are of no use to describe small scales, as the nuclear and even the atomic scale already, it is accepted by the physical community  that   we need to quantize gravity in order to depict our Universe at very early times, at least up to Planck scales (beyond that scale no physical theory has been proven right, up to now). But, for the moment,  nobody knows how to obtain a quantum theory of gravity, it even might be simply impossible. Maybe gravity is a force of a very different nature, as compared with the electromagnetic and the nuclear forces. It could even be non-fundamental, but rather a kind of emergent phenomena not to be described with gauging and quantization. 
 
 In short, adhering to Einstein's viewpoint, we should not worry too much about these mathematical singularities, since nature, and true physical descriptions of it, are always free of them. 
 
 \
 
 In spite of all these problems, some {scientific communities} attempt to obtain physical (or, if you want, metaphysical, until they can be validated with actual experiments) laws valid at high energy densities, in several different ways:
 
 \

\begin{enumerate}
\item
 Introducing quantum effects (more precisely holonomy corrections),  which could be disregarded at low energy densities, as in { Loop Quantum Cosmology} (LQC). These quantum effects produce a modification of the 
FE (the holonomy corrected FE), which becomes \cite{singh0, singh1,singh2,aho}
\begin{eqnarray}\label{LQC}H^2=\frac{\rho}{3M_{pl}^2}\left( 1-\frac{\rho}{\rho_c}  \right),
\end{eqnarray}
where, $\rho_c\cong 0.4 \rho_{pl}$ is the so-called {{\it critical density}}.

\

Note that, now the modified FE depicts an ellipse in the plane $(H,\rho)$ (recall that in GR the FE depicts a parabola), which is a bounded curve, so the energy density is always finite. In fact, it is always bounded by the critical one. As a consequence the singularities such as the BB or the  Type I and III do not exist in LQC. About the BB singularity, people say that, in LQC, the Big Bang is replaced by a Big Bounce (the universe bounces from the contracting to the expanding phase).

\

In addition, for low energy densities $\rho\ll \rho_c$  the FE in GR (the usual one) is recovered. Thus, singularities of Type II and IV also exist in LQC.
Finally, for a non-phantom fluid the movement through all the   ellipse is clockwise, that is, the Universe starts in the contracting phase with an infinite size and bounces to enter in the expanding one.

\

\item  Introducing non-linear effects in the Ricci scalar. We have already seen that in GR the Hilbert-Einstein Lagrangian is ${\mathcal L}=V\left(\frac{R M_{pl}^2}{2}-\rho\right)$. Then, the idea is that at high energy densities the {Ricci} scalar
is replaced by a more general function $R\rightarrow F(R)$, which satisfies $F(R)\rightarrow R$, when $R\rightarrow 0$, in order to recover GR at low energy densities (recall that $R=6(\dot{H}+2H^2)\sim \rho/M_{pl}^2$) \cite{Faraoni, Felice}.

\

To obtain the
Hamiltonian of the system, one has to use the so-called
{Ostrograski} construction, because the Lagrangian contains
second derivatives of $V$. The Hamiltonian constrain leads to the following { modified FE in $F(R)$ gravity}
\begin{eqnarray}
-6F''(R)\dot{R}H+F'(R)(R-6H^2)-F(R)+\frac{2\rho}{M_{pl}^2}=0.
\end{eqnarray}

This is a second-order differential equation with respect to the Hubble rate, 
which together with the CE and the EoS give the dynamics of the Universe. However, contrary to the constituent equations in GR, {one needs to perform numeric calculations in order to solve the equations in $F(R)$-gravity}. The most famous model is  $R^2$-gravity (sometimes known as Starobinsky model)
 given by  $F(R)=R+\alpha R^2$, which was extensively studied in the Russian literature.

\item One can also introduce quantum effects produced by massless fields conformally coupled  with gravity,  obtaining the so-called {\it semiclassical gravity}.
If one considers some  massless fields conformally coupled with gravity, the vacuum  stress-tensor acquires  an anomalous trace, given by \cite{Davies}
\begin{equation}
    T_{vac}= \alpha \partial_{\mu}\partial^{\mu} R-\frac{\beta}{2}G,
\end{equation}
where $R$ is, once again, the Ricci scalar and $G=-2(R_{\mu\nu}R^{\mu\nu}-\frac{1}{3}R^2)$  the Gauss-Bonnet invariant. In terms of the Hubble parameter, one has
\begin{eqnarray}
  T_{vac}=6\alpha\left(\frac{d^3H}{dt^3}+12H^2\dot{H}+7H\ddot{H}+4\dot{H}^2  \right) 
  -12\beta(H^4+H^2\dot{H}).
\end{eqnarray}

The coefficients $\alpha$ and $\beta$ are fixed by the regularization process. For instance, using adiabatic regularization, one gets \cite{Fischetti}
\begin{eqnarray}
    \left. \begin{array}{cc}
       \alpha=  & \frac{1}{2880\pi^2}(N_0+6N_{1/2}+12N_1)>0, \\
       & \\
      \beta=  & \frac{-1}{2880\pi^2}(N_0+\frac{11}{2}N_{1/2}+62N_1)<0,      \end{array}\right.
\end{eqnarray}
while point splitting yields \cite{Davies}
\begin{eqnarray}
    \left. \begin{array}{cc}
       \alpha=  & \frac{1}{2880\pi^2}(N_0+3N_{1/2}-18N_1), \\
       & \\
      \beta=  & \frac{-1}{2880\pi^2}(N_0+\frac{11}{2}N_{1/2}+62N_1),      \end{array}\right.
\end{eqnarray}
where $N_0$ is the number of scalar fields, $N_{1/2}$ that  of four-component neutrinos, and $N_1$ the number of electromagnetic fields. 

\

Here it is important to note, as pointed out in \cite{Wald}, that the coefficient $\alpha$ is arbitrary, although it is influenced by the regularization method and also by the fields present in the Universe, but $\beta$ is independent of the regularization scheme and it is always negative.  

\

Now, we are interested in the value of the vacuum energy density, namely $\rho_{vac}$. Since the trace is given by $T_{vac}=\rho_{vac}-3P_{vac}$, inserting this expression in the conservation equation $\dot{\rho}_{vac}+3H(\rho_{vac}+P_{vac})=0$, one gets
\begin{equation}
    \dot{\rho}_{vac}+4H\rho_{vac}-HT_{vac}=0,
\end{equation}
 a first-order linear differential equation, which can be integrated by using the method of variation of constants, leading to
\begin{equation}
    \rho_{vac}=6\alpha\left(3H^2\dot{H}+H\ddot{H}-
    \frac{1}{2}\dot{H}^2 \right)-3\beta H^4+Ca^{-4},
\end{equation}
where $C$ is an integration constant, which vanishes for  flat space-time. This can be understood as follows: for a static space-time $\rho_{vac}$ reduces to $Ca^{-4}$, and the flat  space-time reduces to  Minkowski, for which $\rho_{vac}=0$, and thus, $C=0$. Therefore, in semi-classical gravity,
the Friedmann equation becomes 
\begin{equation}\label{semiclassical}
    H^2=\frac{\rho+\rho_{vac}}{3M_{pl}^2}.
\end{equation}

\

Here we will consider the empty flat case, which corresponds to $\rho=0$ and $C=0$. There, since $\beta<0$, one has a de Sitter solution $H_+=\sqrt{-\frac{1}{3\beta}}$.
In addition, the Friedmann equation is, in the empty case,
\begin{eqnarray}
3M_{pl}^2H^2= 6\alpha\left(3H^2\dot{H}+H\ddot{H}-
    \frac{1}{2}\dot{H}^2 \right)-3\beta H^4,
\end{eqnarray}
and, for $H=0$, becomes
\begin{eqnarray}
\ddot{H}=\frac{\dot{H}^2}{2H},
\end{eqnarray}
whose solution is given by $H(t)=at^2$; what shows that for this model the branches $H>0$ and $H<0$ decouple, i.e., the universe cannot transit from the expanding to the contracting phase, and vice versa.

\

Next, performing the change of variable $p=\sqrt{H}$ (we are here considering that the universe expands), the semi-classical Friedmann equation  becomes \cite{Wada}
\begin{eqnarray}
\frac{d}{dt}\left(\dot{p}^2/2+V(p)\right)=-p^2\dot{p}^2,
\end{eqnarray}
where
\begin{eqnarray}
V(p)=-\frac{p^2}{24\alpha}\left( 1+\beta p^4 \right).
\end{eqnarray}

The corresponding dynamical system can be written as 
\begin{eqnarray}
  \left\{ \begin{array}{ccc}
   \dot{p}    & = & y\\
    \dot{y}   & = & -3p^2y-V'(p),
  \end{array}\right.
\end{eqnarray}
 which can be simply viewed as the dynamics,  with  friction, of a particle under the action of a potential.
 
 There are two different situations (we use the notation $p_+=\sqrt{H_+}$):  
\begin{enumerate}
    \item  Case $\alpha >0, \beta<0$.
Here the system has two fixed points: $(0,0)$ is an unstable critical point, and $(p_+, 0)$ is stable (it is the minimum of the potential). Solutions
are only singular at early times. At late times they oscillate and shrink around a stable point, that is, $(p_+, 0)$ is a
global attractor. In addition, there is a solution that ends at $(0, 0)$, and only a nonsingular solution that starts at $(0, 0)$ (with zero energy) and ends at $(p_+, 0)$.
\item  Case $ \alpha < 0, \beta < 0$.
This is the famous Starobinsky model \cite{Starobinsky}. The system has two critical points: $(0, 0)$ is a stable critical point, and $(p_+, 0)$
is a saddle point (it is the maximum of the potential). There are solutions that do not cross the axis $p = p_+$; these solutions are singular at early and late
times: they correspond to the trajectories that cannot pass the top of the potential. There are other solutions that cross
 the axis $p = p_+$ twice; they are also singular at early and late times. These trajectories pass the top of the potential
bounce at $p = 0$ and pass once again the top of the potential. There are solutions that cross the axis $p = p_+$ once;
these solutions are singular at early times, however at late times the solutions spiral and shrink to the origin. These
solutions pass the top of the potential once, and then bounce some times about $p = 0$, shrinking to $p = 0$. 
Finally, there are only two unstable
no-singular solutions: one goes from $(p_+, 0)$ to $(0, 0)$, and the other is the de Sitter solution $(p_+, 0)$.   

\

  What seems a little bit strange is that the title of  Starobinsky's paper \cite{Starobinsky} is  "A new type of isotropic cosmological models without singularity". In fact, as we have just shown, in that model there is only one non singular solution, but it is unstable, and thus non-physical.

   \end{enumerate}

\end{enumerate}

 \subsection{Chronology of the Universe} 
 
{To explain the different phases of the Universe, first of all,  we need to recall some basic elements of thermodynamics.}
For a relativistic fluid (made of light particles with velocities comparable to the speed of light) in thermal equilibrium at temperature $T$, we know that:
\begin{enumerate}
\item The energy density is given by  $\rho=\frac{\pi^2}{30}g_*T^4$, where $g_*$ is the number of degrees of freedom, which for the modes in the Standard model are $28 + \frac{7}{8}\times  90 = 106.75$.
\item For a relativistic fluid, pressure is related with energy via the linear relation $P=\rho/3\Longleftrightarrow w_{eff}=1/3$.
\item The number density of particles is $n=\frac{\zeta(3)}{\pi^2}g_* T^3$, where $\zeta$ is the {Riemann} zeta function.
\item The entropy density is  $s=\frac{\rho+P}{T}=\frac{2\pi^2}{45}g_* T^3 $.
\end{enumerate}

\

The total entropy is $S\equiv s a^3$, so for an {\it adiabatic process}, i.e., when the total entropy is conserved, one has 
\begin{eqnarray} T a= \mbox{ constant} \Longrightarrow T \frac{a}{a_0}= \mbox{ constant} 
\Longrightarrow \frac{T}{z+1} = \mbox{ constant}, \end{eqnarray}
where $a_0$ is the present value of the scale factor and $z$  the cosmic red-shift.

\

Since the current temperature of our Universe is $T_0\cong 2.73$ K $\cong 2.3\times 10^{-13}$ GeV, one has
\begin{eqnarray}
\frac{T/\mbox{GeV}}{z+1}=2.3\times 10^{-13}.
\end{eqnarray}

 \
 
 We also need the relation between the temperature of the Universe and its age. To get it, we use the the  value of the Hubble rate for a radiation dominated universe $H(t)=\frac{1}{2t}$. From the FE and the Stefan-Boltzmann law, in natural  units,  we have 
 \begin{eqnarray}\label{temp}
 \frac{1}{4t^2}=\frac{\pi^2}{90}g_*\frac{T^4}{M_{pl}^2}\Longrightarrow T=\sqrt{\frac{M_{pl}}{2\pi}}\left(\frac{90}{g_*} \right)^{1/4}\frac{1}{\sqrt{t}}.
 \end{eqnarray}
 
 \
 
 Next, denoting by $t_{sec}$ the age of the Universe in seconds, we have $$t_{sec}\times 1 s\cong 2 t_{sec} \times 10^{43} t_{pl}\cong 
 4t_{sec} \times 10^{42} M_{pl}^{-1}.$$ 
 
 Inserting this relation in (\ref{temp}), one obtains the following expression for the temperature of the Universe (in MeV) in terms of its age (in seconds):
\begin{eqnarray} T_{MeV}=\frac{\mathcal{O}(1)}{\sqrt{t_{sec}}}.
\end{eqnarray}

\

From there, in the BB model the chronology of our Universe can be described as follows.

\begin{enumerate}
\item  Planck scale.  $T_{pl}\sim M_{pl}\cong 2.4\times 10^{21}$ MeV, what means that the Planck scale is reached at $t_{pl}\sim 10^{-43}$ seconds after the BB. It is very important to remark that no reliable physical theory can be invoked before this time. This is a crucial, insurmountable constraint that any serious physicist knows well, but that, too often, is kept hidden under the carpet (nobody seems to be interested in proclaiming the shortcomings of present day fundamental physics). The corresponding  red-shift has a value of $z_{pl}\sim 10^{31}$

\item   Gran Unification Theory (GUT) scale. It enters when the temperature goes down to  $T_{GUT}\sim 10^{16}$ GeV, i.e., for  $t_{GUT}\sim 10^{-36}$s or $z_{GUT}\sim 10^{29}$. The three forces of the standard model (electromagnetic, weak and strong), which constituted a unique force until then, start to become separated forces below this temperature.

\item Electroweak epoch. It occurs at  $T_{EW}\sim 10^{15}-10^9$ GeV, i.e., when $t_{EW} < 10^{-32}$s or $z_{EW}<10^{22}$. The strong interaction clearly decouples from the electroweak one.

\item Radiation dominated era. It sets up when $T_{rad}\sim 1$ eV, i.e, $t_{rad}\sim 10^{12}$s or $z_{rad}\sim 4000$. The energy density of nearly massless relativistic particles dominates. From then on, the weak and electromagnetic forces become separated so that, finally, during this period all the different forces decouple and become distinct. Electromagnetic radiation dominates the energy content of the universe at this epoch.

\item  Matter domination era.  It occurs for $T_{matt}< 1$ eV.  The energy density of matter dominates, at last. Clusters of galaxies and stars start to be  formed during this period due to the omnipresent gravitational force that now overcomes radiation pressure.

\item Recombination. This period starts when the temperature goes down to around $3,000$ K , the redshift being $z_{rec}\sim 1,100$. This happens some $300,000$ years after the BB. 
At this epoch, nearly all free electrons and protons recombine and form neutral hydrogen.
Photons decoupled from matter and they can travel freely, for the first time, throughout the whole Universe. They originate what is observed today as the Cosmic Microwave Background  (CMB) radiation 
(in that sense, the cosmic background radiation is infrared [and some red] black-body radiation emitted when the universe was at a temperature of some $3,000$ K, redshifted by a factor of 1,100 from the visible spectrum to the microwave spectrum).

\item  Dark Energy era. It starts  when the falling temperature reaches $T_{DE}\sim 10^{-3}$ eV (correspondingly, $z_{DE}\sim 4$), that is, around $3$ billion years ago, and lasts up to present time, extending into the future. The Universe is dominated by an unknown sort of energy, called  {\it dark energy}, and, under its influence, it starts to accelerate  (so-called current cosmic acceleration or late-time acceleration).

\end{enumerate}

\

A final, but important, observation is that, 
as we will discuss in the next Section, the classic BB model, in spite of being able to describe the first stages of the universe evolution, had some very serious shortcomings, which could only be overcome by introducing, at a very early time -most likely  at GUT scales- a new, very brief stage named {\it cosmic inflation}, where the volume of our Universe inflated by more than {$65$} e-folds in an {incredibly small period of time.}

 \section{Inflation}
 
 BB Cosmology had some serious troubles,  the most famous of them being the horizon problem, pointed out for the first time by Wolfgang Rindler, and the flatness issue, clearly described by Robert Dicke. 
 
 \subsection{The problems of BB Cosmology}

\begin{enumerate}
\item {The horizon problem:}

Imagine two observers, $A$ and $B$, in a circumference of variable radius $a(t)$ separated by and angle $d\theta$. The time it takes for a light signal emitted from A to reach B
 is $dt$. Then, as in natural units the speed of light is $1$, we have $a(t)\frac{d\theta}{dt}=1$, the distance travelled by a signal of light emitted at time $t_i$ and arriving at time $t_f$ is $d(t_i,t_f)=a(t_f)\int_{t_i}^{t_f}\frac{dt}{a(t)}$.  This is what happens in an expanding universe. Let $t_0$ be the present time and we take $t=0$ at the singularity, i.e., when the BB occurs. Thus,  the distance travelled by a signal emitted at time $t=0$  and received now is 
$$d_0=a_0\int_{0}^{t_0}\frac{dt}{a(t)}.$$ 

This is the {present horizon size} (the size of our patch of Universe). {We cannot see beyond that distance, i.e., 
we cannot see galaxies which are further away than the present horizon size!}

 \

 To simplify, we will assume that the Universe is matter dominated, that is, $H(t)=\frac{2}{3t}\Longrightarrow a(t)=a_0\left( \frac{t}{t_0} \right)^{2/3}$, and thus,
${d_0}\sim t_0\sim {H_0^{-1}}$. We know that our Universe is now very homogeneous, so at Planck scales it had to  be extremely homogeneous, but at that time the size of our Universe was
${d_{pl}=\frac{a_{pl}}{H_0a_0}}$ (Recall that, due to the Hubble law,   in an expanding Universe ${d_{AB}(t_2)=\frac{a(t_2)}{a(t_1)}d_{AB}(t_1)}$).

\

This quantity has to be compared with the size of the causal regions (the distance that light travels from the Big Bang to the Planck era), which is ${d_c}\sim t_{pl}\sim
{H_{pl}^{-1}}$. We calculate the ratio
\begin{eqnarray}\frac{d_{pl}}{d_c}\sim \frac{H_{pl}a_{pl}}{H_0a_0}={\frac{\dot{a}_{pl}}{\dot{a}_0}}= \frac{H_{pl}T_0}{H_0T_{pl}}   \sim \frac{T_0}{H_0},\end{eqnarray}
where we have used the adiabatic evolution of the Universe, $aT=$ constant, and that, in natural  units, $H_{pl}\sim T_{pl}\sim M_{pl}$.

\

Using now that $H_0\sim 6\times 10^{-61} M_{pl}$ and $T_0\sim 5\times 10^{-31} M_{pl}$, we obtain
$$\frac{d_{pl}}{d_c}\sim {10^{30}},$$
what means that, at Planck scales, there are ${10^{90}}$ disconnected regions. Then, assuming that inhomogeneities cannot be dissolved by ordinary expansion, 
how is it possible that our present Universe be so homogeneous and that the  Cosmic Microwave Background (CMB) radiation has practically the same temperature in all directions? 
This seems impossible, if it comes from a patch, which at Planck scales, contains  so many regions that have never been in causal contact (they have never exchanged information). This is the well-known horizon problem.

\

An equivalent way to see this problem goes as follows:
 As we have already explained, the decoupling, or the last scattering, is thought to have occurred at recombination, i.e., about $300,000$ years after the Big Bang, or at a redshift of about ${z_{rec}\approx 1,100}$. We can determine both  the size of our Universe   and the physical size of the particle horizon that had existed at this time. 
   
   The size of our Universe coincides approximately with the size of the last scattering surface, which nowadays is approximately $H^{-1}_0$, so that, at  recombination,  the diameter of the last scattering surface was $d_{rec}=
   \frac{a_{rec}}{a_0H_0}$. At that time, the size of a causally connected region is $d_c\sim H_{rec}^{-1}$. Then, we have
\begin{eqnarray}
\frac{d_{rec}}{d_c}\sim  \frac{H_{rec}a_{rec}}{H_0a_0}= \frac{H_{rec}}{H_0 (1+z_{rec})}. \end{eqnarray}
Next,  taking into account that the energy density of matter scales as
$\rho_{m,rec}=\rho_{m,0}(1+z_{rec})^3$, further, that at the recombination the universe is matter-dominated, i.e.,
$H_{rec}=\sqrt{\frac{\rho_{m,rec}}{3M_{pl}^2}}$, and that at  present time $\rho_{m,0}=3\Omega_{m,0}M_{pl}^2 H_0^2$, with $\Omega_{m,0}\cong 0.3$, one obtains
\begin{eqnarray}
\frac{d_{rec}}{d_c}\sim  \sqrt{\Omega_{m,0}(1+z_{rec})}\sim 18.
\end{eqnarray}   
 As a consequence, in the last scattering surface there are regions which are causally disconnected; but it turns out that the CMB has practically the same temperature in all directions.

   \
   
 To simplify, the horizon size is of the order $1/H_0\sim$ 14 billions of light-years, which coincides with the age of the Universe $1/H_0\sim $ 14 billions of years. So, imagine a region that is at a distance of 10 billions of light-year  from us, and other region, in the opposite direction, that is at the same distance form us. The question is, how it is possible that both regions, which are about 20 billion of light-years apart, they emitted light at the same temperature?
 
 This is an apparent "paradox" in an static or decelerating universe, but as we will see in next Section,  the paradox is overcome when one assumes a short superluminal expansion phase at early times.

\item The flatness problem:

 Up to now, we have only considered the dynamical equation for  flat space, but in fact space could have positive or negative curvature. When one considers
 the general case, the FE  become
 \begin{eqnarray} \label{FEcurvature}{H^2=\frac{\rho}{3M_{pl}^2}-\frac{\kappa}{a^2}},\end{eqnarray}
 with $\kappa=-1,0,1$ (open, flat and closed cases, respectively).
 
 This equation can be written as follows
 $$\Omega-1=\frac{\kappa}{a^2 H^2},$$
 where ${\Omega\equiv \frac{\rho}{3H^2M_{pl}^2}}$ is the ratio of the energy density to the {{\it critical one}}.

 \
 
 Evaluating at the Planck time and at the present time, one gets:
  \begin{eqnarray}
 \frac{|\Omega-1|_{pl}}{|\Omega-1|_0}=\frac{H_0^2a_0^2}{H_{pl}^2a_{pl}^2}={
 \frac{\dot{a}_0^2}{\dot{a}_{pl}^2}}=\frac{H_0^2T_{pl}^2}{H_{pl}^2T_0^2}\sim \frac{H_0^2}{T_0^2}\sim 10^{-60},
 \end{eqnarray}
 where we have used the adiabatic evolution of the universe.

 Thus, in order to have $|\Omega-1|_0\cong 0.01$ (the present observational data), the value of $\Omega-1$ at early times has to be fine-tuned to values amazingly close to zero, but without letting it be exactly zero! This is the flatness problem which sometimes is also dubbed as {\it the fine-tuning problem}.
 
 \
 
 To better understand this fine tune: 
 Our very existence depends on the fanatically close balance between the actual density and the critical density in the early universe. If, for instance, the deviation of $\Lambda$ from one at the time of Nucleosynthesis  had been one part in $30$ thousand instead of one part in $30$ trillion, the universe would have collapsed in a Big Crunch  after only a few years. In that case, galaxies, stars and  planets  would not have had time to form, and so, cosmologist would never have existed.

 \end{enumerate}

 \subsection{Inflation: the basic idea}
 The solution of the horizon, flatness,  and of the other problems Big Bang cosmology had accumulated during several decades, was obtained by {Alan Guth} in 1981, in his seminal paper:
"{\it The Inflationary Universe: A Possible Solution to the Horizon and Flatness Problems}" (PRD23, pag. 347-356).

\

The key point to solve these problems is to consider the ratio { $\frac{\dot{a}_{pl}}{\dot{a}_0}$} (this quantity appears explicitly in the horizon and flatness problems). If gravity is attractive, then this ratio is necessarily larger than one, because
gravity decelerates the expansion. Therefore, the conclusion
 $\frac{\dot{a}_{pl}}{\dot{a}_0}\gg 1$ can
be avoided only if {we assume that during some period in the cosmic expansion gravity acted
as a "repulsive" force, thus accelerating expansion}. In this case, we can have $\frac{\dot{a}_{pl}}{\dot{a}_0}< 1$  and the creation of our patch of Universe from a single causally connected domain may become possible.

 \
 
 More precisely, 
let $a_b$ and $a_e$ be, respectively, the values of the scale factor at the beginning and at the end of the accelerated expansion period, which is named {\it {inflation}}. 
Since $H=\frac{\dot{a}}{a}$, integrating this equation, we have 
$a_e=a_be^N$, where $N=\int_{t_b}^{t_{e}}H(t)dt$ is {{\it the number of e-folds inflation lasts}}.
Assuming that at the beginning of  inflation, which occurred almost at the very beginning of the Universe, there was an small  patch of it causally connected (light had enough time to travel from one to any other point in that domain), then inflation blows up this region to a very large one, preserving the homogeneity intact while expanding enormously our patch of the Universe. 

 \
 
 In order that this patch encompasses our whole Universe, as we observe it now, we will see that it is necessary that {the number $N$ of e-folds the patch increases has to be higher than $65$}. During inflation {the volume grows}  { $e^{3N}$ times}, (by this proportion, {the volume of an atom would turn into that  of an orange.}) 

 \

Going a bit further, to compute the number of e-folds needed and the time inflation should last, we will assume
 that inflation starts and ends at the GUT scale, i.e, when $H_{GUT}\sim 10^{14}$ GeV and $T_{GUT}\sim 10^{16}$ GeV. 
The size of our Universe at the beginning of inflation is ${ d_{GUT}=\frac{a_b}{H_0a_0}}$ and a causally connected region has size $d_b\sim H^{-1}_{GUT}$. Then,
$$\frac{d_{GUT}}{d_b}\sim \frac{H_{GUT}T_0}{H_0T_{GUT}}\sim 10^{-2}\frac{T_0}{H_0}\sim 10^{28}.$$
Since at the end of inflation the size of a causally connected region is ${d_e=\frac{a_e}{a_b}d_b=e^N d_b}$, we will have 
$$\frac{d_{GUT}}{d_e}\sim e^{-N}10^{28}\leq 1.$$

Thus, 
$$e^N\geq 10^{28}\Longrightarrow {N\geq} 28\ln(10)\sim { 65},$$
as already advanced.

 \
 
 In the same way, for the flatness problem one has 
 \begin{eqnarray}
 |\Omega-1|_0\sim 10^{56} |\Omega-1|_{GUT},
 \end{eqnarray}
 but assuming an inflationary period at GUT scales one has 
 \begin{eqnarray}
 |\Omega-1|_e\sim e^{-2N} |\Omega-1|_{b}.
 \end{eqnarray}
 
 So, supposing that prior to inflation, the universe was actually fairly strongly
$|\Omega-1|_{b}\sim 1$, one has 
 \begin{eqnarray}
 |\Omega-1|_0\sim 10^{56} e^{-2N} |\Omega-1|_{e},
 \end{eqnarray}
 and thus, for $N\geq 65$ the flatness problem is solved.

 \
 
 On the other hand, since during inflation $H$ is {nearly constant} (as we will see in the next subsection),  we  have (approximately)
$N\sim H_{GUT}(t_e-t_b)$.  Then, for $N\sim 10^2$ one gets
$$
t_e-t_b\sim 10^{-12} \mbox{GeV}^{-1}\sim 10^6 M_{pl}^{-1}\sim 10^{-38} \mbox{s}.
$$

\

The following important remark is in order.  By the end of inflation the size of the Universe has {grown extraordinarily}, what means that the particles existing before  are now very diluted, and thus,
the Universe becomes {extremely cold} and has a {very low entropy}. As a consequence, in order to match it with the initial stage of the hot BB model, 
a {{\it reheating mechanism is needed.}} This mechanism is a very non-adiabatic process, in which an enormous amount of particles (practically the whole matter content of the universe, or primordial quark-gluon plasma) is created via  {quantum effects}; after its thermalization, the Universe also becomes reheated, and eventually matches with the starting conditions of the hot BB Universe.

 \

 \subsection{A simple way to produce inflation}

We  consider a  Universe filled out with an homogeneous scalar field, namely $\phi(t)$, whose potential is $V(\phi)$. To start, we assume a static Universe, and thus the dynamical equation of a field can be readily obtained from the first law of thermodynamics:
\begin{eqnarray}\label{law}
d(\rho a^3)=-Pd(a^3).
\end{eqnarray}
Since we are assuming a static universe, $d(a^3)=0$, and taking into account that the energy density is equal to the sum of its kinetic plus its potential part, i.e., 
$\rho=\frac{\dot{\phi}^2}{2}+V(\phi)$, the first law becomes
\begin{eqnarray}
d(\rho)=0 \Longrightarrow \ddot{\phi}+V_{\phi}(\phi)=0,
\end{eqnarray}
which is the same equation as the one for a particle under the action of a potential that can be obtained from the Lagrangian
\begin{eqnarray} \label{L} {\mathcal L}=\left(\frac{\dot{\phi}^2}{2}-V(\phi)\right) a^3,\end{eqnarray}
by using the Euler-Lagrange equations.

\

We now  consider an expanding universe, where, from the Lagrangian  (\ref{L}), we obtain the dynamical equation
\begin{eqnarray}\label{CEfield}\ddot{\phi}+3H\dot{\phi}+V_{\phi}(\phi)=0,
\end{eqnarray}
which could also be derived from the first law of thermodynamics (\ref{law}): 
\begin{eqnarray}
\left( \ddot{\phi}+V_{\phi}(\phi)  \right)\dot{\phi}a^3+
3H\left(\frac{\dot{\phi}^2}{2}+V(\phi)\right)a^3=-3HPa^3.
\end{eqnarray}
Comparing  this expression with the dynamical equation 
(\ref{CEfield}), we conclude that the pressure when the universe is filled by an scalar field is 
\begin{eqnarray}\label{field}
 P=\frac{\dot{\phi}^2}{2}-V(\phi).\end{eqnarray}

 \
 
 Coming back to the dynamics, we have
 an autonomous  second-order differential equation
\begin{eqnarray}\ddot{\phi}+3H\dot{\phi}+V_{\phi}(\phi)=0,
\end{eqnarray}
where $H=\frac{1}{\sqrt{3} M_{pl}}\sqrt{\frac{\dot{\phi}^2}{2}+V(\phi)}$,
which
unfortunately, it is impossible to find analytic solutions as in the case of a fluid. {In fact, the system can only be solved numerically}.

\

Recalling now the  equation for the acceleration $
\frac{\ddot{a}}{a}=\dot{H}+H^2=-\frac{{H}^2}{2}(1+3w_{eff}),
$ where $w_{eff}=P/\rho=-1-\frac{2\dot{H}}{3H^2}$, and thus,  the condition for an accelerated expansion is {$\rho+3P<0$ or $w_{eff}<-1/3$}. In a similar way,  in terms of the Hubble rate and its derivative, we  also see that the condition for acceleration is ${-\frac{\dot{H}}{H^2}<1}$, and that acceleration ends when
$-\frac{\dot{H}}{H^2}=1$.

\

So, a necessary condition to have accelerated expansion is { $\dot{\phi}^2<V(\phi)$}, because,
\begin{eqnarray}
\rho+3P<0\Longrightarrow \dot{\phi}^2<V(\phi).
\end{eqnarray}

\

Here it is important to realize that the condition ${P\cong -\rho}$, and thus ${\dot{H}\ll H^2} $, is enough to have acceleration. Since the condition $P\cong -\rho$ is equivalent to $\dot{\phi}^2\ll V(\phi)$,  the successful realization
of inflation  requires keeping $\dot{\phi}^2$ small as compared with $V(\phi)$ (the kinetic energy must be small compared with the potential one) during a sufficiently
long time interval; more precisely, for at least, $65$ e-folds, but this will actually depend on the shape of the potential.
{In practice one needs a very, very flat potential}.

\

Then, with this condition, the FE becomes $H^2\cong\frac{V(\phi)}{3M_{pl}^2}$. We also assume, during inflation,  the condition $\ddot{\phi}\ll 3H\dot{\phi}$,  and thus, 
the CE  becomes $3H\dot{\phi}+V_{\phi}(\phi)\cong 0$. Note that for a flat potential one has $V_{\phi}(\phi)\cong 0\Longrightarrow \dot{\phi}\cong 0$, and the resulting Hubble rate is nearly constant during this period, what implies a nearly  exponential grow of the volume of the Universe during this epoch. This is exactly what we need to solve both the horizon and the flatness problems.

\

When both conditions, 
\begin{eqnarray}\label{slow roll conditions}
{\dot{\phi}^2\ll V(\phi)}\qquad \mbox{and}\qquad { \ddot{\phi}\ll 3H\dot{\phi}},
\end{eqnarray}
are fulfilled, we say that the Universe is in the {{\it slow roll regime}}, 
which is mathematically an {attractor} (see, for instance, Section 3.7 of [3]), and the dynamical equations become
\begin{eqnarray}\label{slow roll equations}
H^2\cong\frac{V(\phi)}{3M_{pl}^2} \qquad \mbox{and}\qquad 3H\dot{\phi}+V_{\phi}(\phi)\cong 0.
\end{eqnarray}

\

As an exercise,  we now show  that a necessary  condition to have a slow roll regime is that the {{\it slow roll parameters}} satisfy
\begin{eqnarray}\label{slowrollparameters}{ \epsilon\equiv \frac{M_{pl}^2}{2}\left( \frac{V_{\phi}}{V} \right)^2}\ll 1 \quad \mbox{and} \quad  { \eta\equiv M_{pl}^2\left|\frac{V_{\phi\phi}}{V} \right|}\ll 1.\end{eqnarray}
 In fact, from (\ref{slow roll equations}) one has
 \begin{eqnarray}
 \frac{V_{\phi}}{3H}\cong -\dot{\phi}\Longrightarrow 
 \frac{V^2_{\phi}}{9H^2}\cong \dot{\phi}^2\ll V\Longrightarrow 
 \frac{M_{pl}^2}{2}\frac{V_{\phi}^2}{V^2}\ll \frac{3}{2}\sim 1\Longrightarrow \epsilon\ll 1,
 \end{eqnarray}
 where we have used that during the slow roll regime $H^2\cong \frac{V}{3M_{pl}^2}$. Finally, to obtain the condition $\eta\ll 1$, we start taking the temporal derivative of the slow roll equation
 $3H\dot{\phi}+V_{\phi}(\phi)\cong 0$, to obtain
 \begin{eqnarray}
 \frac{V_{\phi\phi}}{V}\cong -\frac{3\dot{H}}{V}-\frac{3H\ddot{\phi}}{V\dot{\phi}}\Longrightarrow
 M_{pl}^2\left|\frac{V_{\phi\phi}}{V} \right|\ll
 \frac{3|\dot{H}|M_{pl}^2}{V}+3,
 \end{eqnarray}
 where we have used the slow roll condition $\ddot{\phi}\ll 3H\dot{\phi}$ and the Friedmann equation in the slow roll regime. And, from the Raychaudury equation, $\dot{H}=-\frac{\dot{\phi}^2}{2M_{pl}^2}$, we get
 \begin{eqnarray}
  M_{pl}^2\left|\frac{V_{\phi\phi}}{V} \right|\ll \frac{3\dot{\phi}^2}{2V}+3\ll \frac{9}{2}\sim 1\Longrightarrow \eta\ll 1,
 \end{eqnarray}
 where we have used that, during the slow roll regime, the kinetic energy is smaller than the potential one, i.e., $\dot{\phi}^2\ll V$.
 
 \

 \
 
At this point, we want to understand a bit better 
the meaning of the slow roll conditions. The first condition, ${\dot{\phi}^2\ll V(\phi)}$, is clear enough for producing acceleration, because it implies $w_{eff}\cong -1$. To understand the other one, ${\ddot{\phi}\ll 3H\dot{\phi}}$, we consider an harmonic oscillator with friction: 
\begin{eqnarray}
{\ddot{x}+\gamma \dot{x}+\omega^2x=0}, \qquad \mbox{with} \qquad \gamma>0,
\end{eqnarray}
which  corresponds to the  movement of a particle under the influence of the potential $V(x)=\omega^2 x^2/2$.

\

The general solution is given by 
\begin{eqnarray}
{ x(t)=C_+ e^{\lambda_+ t}+C_- e^{\lambda_- t}},\end{eqnarray}
where
\begin{eqnarray}\lambda_{\pm}=\frac{-\gamma\pm \sqrt{\gamma^2-4\omega^2}}{2}.
\end{eqnarray}
 
 \

Observe that the friction term ${\gamma \dot{x}}$ damps the velocity of the particle. In addition, for a very flat potential, i.e., $\omega^2\ll \gamma^2$, one obtains 
\begin{eqnarray}
{\lambda_+\cong -\omega^2/\gamma} \qquad \mbox{and} {\qquad \lambda_-\cong -\gamma.}
\end{eqnarray}
So, at late time the solution  becomes $x(t)\cong C_+e^{-\omega^2 t/\gamma}$, what means that the solution of the following equation (the slow-roll solution)
\begin{eqnarray}
{ \gamma \dot{x}+\omega^2 x=0},
\end{eqnarray}
which  is given by $e^{-\omega^2 t/\gamma}$, is {an attractor.} 
 
 \
 
 Having now better grasped the slow-roll conditions,
 we introduce another pair of  slow-roll parameters, namely
\begin{eqnarray}
{ \bar{\epsilon}\equiv -\frac{\dot{H}}{H^2}}, 
\qquad { \bar{\eta}\equiv 2\bar{\epsilon}-\frac{\dot{\bar\epsilon}}{2H\bar{\epsilon}}}.\end{eqnarray}

\

It is not difficult to see that, 
 using the {Friedmann} and the {Raychaudury} equations,  the condition ${\bar{\epsilon}\ll 1}$ implies ${\dot{\phi}^2\ll V(\phi)}$ (first  slow-roll condition). In fact, 
 \begin{eqnarray}
 -\dot{H}\ll H^2\Longrightarrow \frac{\dot{\phi}^2}{2}\ll \frac{\dot{\phi}^2+2V}{6}\Longrightarrow  \dot{\phi}^2\ll V(\phi),
 \end{eqnarray}
 where we have used the Friedmann and the Raychaudury equations.
 
 \

On the other hand, the condition ${\bar{\eta}\ll 1}$ together with ${\bar{\epsilon}\ll 1}$ implies 
${ \dot{\bar\epsilon}\ll 2H\bar{\epsilon}}$.
Next, we make use of
\begin{eqnarray}
\dot{\bar\epsilon}=-\frac{\ddot{H}}{H^2}+2H\bar{\epsilon}^2
\end{eqnarray}
in order to obtain
\begin{eqnarray}
-\frac{\ddot{H}}{H^2}\ll 2H(\bar{\epsilon}-\bar{\epsilon}^2)\sim 2H\bar{\epsilon},\qquad  \mbox{because} \qquad \bar{\epsilon}^2\ll
\bar{\epsilon}.
\end{eqnarray}

 \

 Using the definition of $\bar{\epsilon}$, the last condition is equivalent to
\begin{eqnarray}
-\ddot{H}\ll -2H\dot{H},
\end{eqnarray}
and taking into account  the {RE}, ${\dot{H}=-\frac{\dot{\phi}^2}{2M_{pl}^2} \Longrightarrow \ddot{H}=-\frac{\ddot{\phi}\dot{\phi}}{M_{pl}^2}}$, we get
\begin{eqnarray}
\ddot{\phi}\ll H\dot{\phi}\sim 3H\dot{\phi} \qquad \mbox {(second slow-roll condition)}.
\end{eqnarray}

In this way, we have proved that  sufficient conditions to have the slow-roll regime are { $\bar{\epsilon}\ll 1$} and 
{ $\bar{\eta} \ll 1$}. 

\

{ Finally we show that during the slow-roll regime one has {$\bar{\epsilon}\cong\epsilon$} and 
{$\bar{\eta}\cong\eta$}. Effectively, during slow-roll we have
\begin{eqnarray}
\bar{\epsilon}\cong\frac{3\dot{\phi}^2}{2V}\cong\frac{V_{\phi}^2}{6H^2V}\cong\epsilon.
\end{eqnarray}
On the other hand, a simple calculation leads to
\begin{eqnarray}
\bar{\eta}=\bar{\epsilon}-\frac{\ddot{H}}{2H\dot{H}}\cong {\epsilon}-\frac{\ddot{H}}{2H\dot{H}}.
\end{eqnarray}
Next, we have
\begin{eqnarray}
\frac{\ddot{H}}{2H\dot{H}}\cong -\frac{1}{2H\dot{H}}\left( -\frac{V_{\phi\phi}\dot{\phi}^2}{3HM_{pl}^2}+\frac{V_{\phi}\dot{H}\dot{\phi}}
{3H^2M_{pl}^2} \right)\cong -\frac{V_{\phi\phi}}{3H}-\frac{V_{\phi}\dot{\phi}}{6H^2M_{pl}^2}\cong -\eta+\epsilon,
\end{eqnarray}
where we have used both slow-roll equations.

\

Thus,  in practice, the slow-roll regime is equivalent to {$\epsilon\ll 1$} and {$\eta\ll 1$}.

 \subsection{On the number of e-folds}

 We will now obtain the minimum number of necessary e-folds during the slow-roll regime. Since $H=\frac{d\ln a}{dt}$, from the CE we have
$$
H=\frac{d\ln a}{dt}=\dot{\phi}\frac{d\ln a}{d\phi}=-\frac{V_{\phi}}{3H}\frac{d\ln a}{d\phi} \Longleftrightarrow 3H^2= -{V_{\phi}}\frac{d\ln a}{d\phi}, $$
and now, using the FE, one gets
\begin{eqnarray}\frac{d\ln a}{d\phi}= \frac{1}{\sqrt{2\epsilon} M_{pl}},\end{eqnarray}
 whose solution is
 \begin{eqnarray}\label{efolds}
 a_e=a_b \mbox{exp}\left({\frac{1}{M_{pl}}
 \left|\int_{\phi_b}^{\phi_e}\frac{d\phi}{\sqrt{2\epsilon}}\right|}\right) \Longrightarrow {N=\frac{1}{M_{pl}}
 \left|\int_{\phi_b}^{\phi_e}\frac{d\phi}{\sqrt{2\epsilon}}\right|}. \end{eqnarray}
 
 In addition, since in the slow-roll regime one has
 $\epsilon\cong -\frac{\dot{H}}{H^2}$, from the acceleration equation $\frac{\dot{a}}{a}=\dot{H}+H^2$ one can conclude that inflation ends when $\epsilon=1$.

 \
 
 As an example, we consider the quadratic potential  $V(\phi)=\frac{1}{2}m^2\phi^2$, where a simple calculation leads to
\begin{eqnarray}\epsilon=\eta=\frac{2M_{pl}^2}{\phi^2};
\end{eqnarray}
so,  the slow roll regime is satisfied when $\phi> \sqrt{2}M_{pl}$, and inflation ends at ${\phi_e=\sqrt{2}M_{pl}}$.

\

On the other hand,
$$
N=\frac{1}{M_{pl}}\int_{\phi_e}^{\phi_b}\frac{\phi}{2M_{pl}}d\phi=\frac{1}{4M_{pl}^2}(\phi^2_b-2M_{pl}^2),
$$
what means that, to obtain more than $65$ e-folds, the beginning of inflation has to occur at ${\phi_b>\sqrt{262} M_{pl}}$.

\

\

To finish this section,  it is also important to calculate the last number of e-folds, that is, the number of them from the horizon crossing time to the end of inflation. 

\

The horizon crossing refers to the moment when the "pivot scale", namely $k_*$ in comoving coordinates,  leaves the Hubble horizon; that is,  when $k_*=a_*H_*\Longleftrightarrow\frac{1}{H_*}=\frac{a_*}{k_*}$, where the value of the Hubble rate at this moment is called the {\it  scale of inflation}. 

\

The physical value of the "pivot scale" at present time is usually chosen as (see for instance \cite{planck})
$k_{phys}(t_0)=\frac{k_*}{a_0}= 0.05 \mbox{ Mpc}^{-1}\cong  10^{-58}M_{pl}\sim 10^2 H_0$, where we use the value $H_0\cong 6\times 10^{-61} M_{pl}$.

\ 

To calculate the e-folds from the horizon crossing moment to the end of inflation, we start with the relation $a_{END}=e^{N_*}a_*=e^{N_*}\frac{k_*}{H_*}$, which can be written as follows:
\begin{eqnarray}
\frac{k_*}{a_0H_0}=e^{-N_*}\frac{H_*a_{END}}{a_0H_0}
\Longrightarrow \frac{k_*}{a_0H_0}=e^{-N_*}\frac{H_*}{H_0}\frac{a_{END}}{a_{rad}}\frac{a_{rad}}{a_{matt}}\frac{a_{matt}}{a_0},
\end{eqnarray}
where $rad$ (resp. $matt$) denotes the beginning of radiation (resp. of matter domination, i.e, at the matter-radiation equality).

\

To simplify, we will assume that from the end of inflation to the beginning of kination the EoS parameter $w$ is constant. Then, we have
\begin{eqnarray}
\left( \frac{a_{END}}{a_{rad}} \right)^{3(1+w)}=\frac{\rho_{rad}}{\rho_{END}}, \qquad \left( \frac{a_{rad}}{a_{matt}} \right)^{4}=\frac{\rho_{matt}}{\rho_{rad}}.
\end{eqnarray}
Therefore
\begin{eqnarray}\label{star}
N_*=-5.52+\ln\left( \frac{H_*}{H_0}\right)+\frac{1}{4}\ln\left(\frac{\rho_{matt}}{\rho_{rad}} \right)+\frac{1}{3(1+w)}
\ln\left( \frac{\rho_{rad}}{\rho_{END}} \right)
+\ln\left( \frac{a_{matt}}{a_0}\right).
\end{eqnarray}

\

To perform the calculations, we consider a power law potential of the form $V(\phi)=V_0\left(\frac{\phi}{M_{pl}} \right)^{2n}$. Using the virial theorem, one can show  that, after the end of inflation, when the inflaton oscillates the effective EoS parameter is given by $w=\frac{n-1}{n+1}$.

Now, we need the spectral index of scalar perturbations, defined by $n_s=1-6\epsilon_*+2\eta_*$ and also the ratio of tensor to scalar perturbations $r=16\epsilon_*$, where, once again, the star denotes that the quantities are evaluated at horizon crossing. As a simple exercise, it can be shown that, for our power-law potential, the slow-roll parameter $\epsilon_*$ and the spectral index are related by 
\begin{eqnarray}
\epsilon_*=\frac{n(1-n_s)}{2(n+1)}.
\end{eqnarray}

\

From the slow-roll parameter $\epsilon$ one can calculate the value of the energy density at the end inflation; imposing that inflation ends when $\epsilon=1$, we get
\begin{eqnarray}
\epsilon=2n^2\left( \frac{M_{pl}}{\phi}\right)^2\Longrightarrow \phi_{END}=\sqrt{2}nM_{pl}\Longrightarrow
V(\phi_{END})=2^nn^{2n}V_0.
\end{eqnarray}
And taking into account that inflation ends when
\begin{eqnarray}
w_{eff}=-\frac{1}{3}\Longrightarrow \dot{\phi}_{END}^2=V(\phi_{END})\Longrightarrow
\rho_{END}=\frac{3}{2}V(\phi_{END})=3\times 2^{n-1}n^{2n}V_0.
\end{eqnarray}

\

We also need the power spectrum of scalar perturbations, defined as
\begin{eqnarray}
{\mathcal P}_{\zeta}=\frac{H_*^2}{8\pi^2 M_{pl}^2\epsilon_*}\sim 2\times 10^{-9}\Longleftrightarrow
H_*^2=16\pi^2\times 10^{-9}\epsilon_*M_{pl}^2,
\end{eqnarray}
which is essential to calculate the value of $V_0$. Indeed, the square of the scale of inflation is given by
\begin{eqnarray}
H_*^2=\frac{V(\phi_*)}{3M_{pl}^2}=\frac{V_0}{3M_{pl}^2}\left( \frac{2}{\epsilon_*}\right)^n n^{2n},
\end{eqnarray}
where we have used that, at horizon crossing, $\phi_*=\sqrt{\frac{2}{\epsilon_*}}nM_{pl}$. And inserting the value of the square of the scale of inflation into the formula of the power spectrum, we finally obtain 
\begin{eqnarray}
V_0=\frac{96\pi^2}{n^{2n}}\times 10^{-9}\left(\frac{\epsilon_*}{2} \right)^{n+1}M_{pl}^4\Longrightarrow 
\rho_{END}=72\pi^2\times 10^{-9}\epsilon_*^{n+1}M_{pl}^4.
\end{eqnarray}

Next, we use the Stefan-Boltzmann law at the beginning of radiation and at the matter-radiation equality
\begin{eqnarray}
\rho_{rad}=\frac{\pi^2}{30}g_{rad}T_{rh}^4,\qquad 
\rho_{matt}=\frac{\pi^2}{15}g_{matt}T_{matt}^4,
\end{eqnarray}
where $g_{rad}=106.75$ is the number of effective degrees of freedom for the Standard Model, $g_{matt}=3.36$ the number of the effective degrees of freedom at matter-radiation equality, $T_{rh}$ the reheating temperature (the temperature of the universe at the beginning of the radiation era), and $T_{matt}$ the temperature of the universe at matter-radiation equality. (Note that, at matter-radiation equality, the energy density of radiation is the same as that for matter; for this reason, the factor $1/15$ appears in the energy density at that moment).

\

Finally, we need use the adiabatic evolution of the universe after the matter-radiation equality, $a_0T_0=a_{matt}T_{matt}$. Inserting all these quantities in (\ref{star}), one gets
\begin{eqnarray}
N_*\cong -15-\frac{5}{1+w}+\left( \frac{1}{2}-\frac{n+1}{3(1+w)} \right)\ln\epsilon_*
+\ln\left( \frac{T_0 M_{pl}}{T_{rh}H_0} \right)+
\frac{4}{3(1+w)}\ln\left(\frac{T_{rh}}{M_{pl}} \right),
\end{eqnarray}
and using that $H_0\sim 6\times 10^{-61}M_{pl}$ and $T_0\cong 2.73 \mbox{ K}\sim 10^{-31} M_{pl}$, the number of  e-folds finally becomes
\begin{eqnarray}
N_*\cong 52-\frac{5}{1+w}+\left( \frac{1}{2}-\frac{n+1}{3(1+w)} \right)\ln\epsilon_*
+\left(1-
\frac{4}{3(1+w)}\right)\ln\left( \frac{ M_{pl}}{T_{rh}} \right).
\end{eqnarray}

\

To finish, inserting the value of $w=\frac{n-1}{n+1}$,
$\epsilon_*=\frac{n(1-n_s)}{2(n+1)}$ and taking the central value of the spectral index $n_s\cong 0.96$, one obtains the last number of e-folds as a function of $n$
\begin{eqnarray}
N_*(n)\cong 52+\frac{n^2-6n-4}{2n}-\frac{n^2-n+1}{6n}\ln\left(\frac{n}{2(n+1)} \right)
+\frac{n-2}{3n}\ln\left( \frac{ M_{pl}}{T_{rh}} \right).
\end{eqnarray}

\

So, for a quadratic potential the number of e-folds as a  function of the reheating temperature is 
$N_*(1)\cong 48-\frac{2}{3}\ln\left( \frac{ M_{pl}}{T_{rh}} \right)$ and, for a quartic potential, one has
$N_*(2)\cong 49$, which does not depend on the reheating temperature, because for a quartic potential, when the inflaton starts to oscillate, $w=1/3$, that is, the universe enters in the radiation phase.

\subsection{Different reheating mechanisms}

\begin{enumerate}
\item In the case that the  potential  has a deep well, at the end of inflation the inflaton field starts to oscillate in this deep well, and releases its energy by creating particles \cite{kofman, kolb,kolb1}. This happens in standard inflation, but after the discover of the current cosmic acceleration, other models containing monotonic potentials appeared, and thus, since the inflaton field cannot oscillate in this case, other mechanisms to reheat the Universe were proposed. 
\item When the potential is a monotonous  function, particles could be created via the so-called {{\it instant preheating}} developed by Felder, Kofman and Linde
\cite{Felder}. In that case, a  quantum scalar field with a very light bare mass is coupled with the inflaton, the adiabatic regime breaks after the end of inflation an particles with a very effective heavy mass are copiously created. The energy density of these particles could never dominate the one of the background, because in that case, another undesirable inflationary period would appear; this is the reason why these particles have to decay in lighter ones well before  they can dominate. Once the dacy is finished, the Universe becomes reheated, at a temperature close to $10^9$ GeV, thus matching with the hot BB model.
\item For a monotonous potential, containing  and abrupt phase transition from the end of inflation to a regime where all the energy density is kinetic (named {\it kination phase} \cite{Joyce} or deflationary phase in  \cite{Spokoiny}),  superheavy  particles \cite{Hashiba,J,ema, hashiba} and also lighter ones \cite{Parker,Zeldovich,ford,gmm,glm,fmm, Damour, Giovannini,Birrell1} can  be created,  via { {\it gravitational particle production}}. The problem of the reheating via production of light particles is that it could appear undesirable  polarization effects which disturb the evolution of the inflaton field during the slow roll period (see for a detailed explanation \cite{Felder}). On the contrary, these polarization effects, during inflation, can be neglected when on consider the production of superheavy particles, which  have to decay in lighter ones, to get, after thermalization, a hot radiation dominated Universe.
\item The { {\it curvaton} mechanism}. Besides the inflaton, there is another massive field, named the {\it curvaton}, which becomes sterile during inflation and at the end of this period, this curvaton field, whose potential has a deep well,  starts to oscillating decaying  into lighter particles \cite{Urena, Feng, Agarwal, Campuzano,Bueno}.
\end{enumerate}
To finish this section, we should add that the reheating parameters (especially the reheating equation of state  parameter) are not sufficiently well constrained. Among other contributions, some possible ways to constrain the reheating EoS parameter have recently been proposed, involving magnetogenesis or primordial gravitational waves. The reheating era has been argued to have a considerable effect on the primordial magnetic field as well as on primordial GWs, which in turn help to extract some viable constraints on the reheating parameters. This has been carried out for two different reheating mechanisms. In the case of GWs, it has been also shown that a late reheating phase helps to improve the fit of the NANOGrav observational data  \cite{tanm1}.

\section{The current cosmic acceleration}

\subsection{The cosmological constant}

The {cosmological constant} (CC), $\Lambda$,  was introduced by {Albert Einstein} (1917) in order to obtain a static model for our Universe (at that time, because of very reasonable physical considerations,  everybody believed that the Universe was static}; see all details in, e.g., \cite{eliz21}). The introduction of $\Lambda$ modifies the FE, as follows
\begin{eqnarray}H^2=\frac{\rho}{3M_{pl}^2}-\frac{\kappa}{a^2}+\frac{\Lambda}{3}.
\end{eqnarray}

From this equation we see at once that the introduction of $\Lambda$ is equivalent to add a new component to the energy content of the Universe, with an energy density 
$\rho_{\Lambda}=\Lambda M_{pl}^2$, and, from the CE, with pressure $P_{\Lambda}=-\rho_{\Lambda}$. Thus, a cosmological constant with positive sign acts against gravitation: ($w_{\Lambda}=P_{\Lambda}/\rho_{\Lambda}=-1$).

\

{Einstein} also considered the RE
\begin{eqnarray}
\dot{H}=-\frac{\rho+P}{2M_{pl}^2}+\frac{\kappa}{a^2}.\end{eqnarray}

Then, for a  matter dominated Universe,  $P=0$, an static solution, $H=\dot{H}=0$, as the one Einstein was looking for,  must satisfy
\begin{eqnarray}\frac{\rho}{3M_{pl}^2}-\frac{\kappa}{a^2}+\frac{\Lambda}{3}=0 \quad \mbox{ and } \quad \frac{\rho}{2M_{pl}^2}=\frac{\kappa}{a^2}.
\end{eqnarray}

From these equations we  see that a static Universe has to be closed, i.e., $\kappa=1$. In Einstein's static model, the energy density and the radius of the Universe are given by
\begin{eqnarray}\rho=2\Lambda M_{pl}^2 \quad \mbox{ and } \quad a=\Lambda^{-1/2},\end{eqnarray}
respectively.

\

Unfortunately,  from  the acceleration equation $\frac{\ddot{a}}{a}=-\frac{\rho}{6M_{pl}^2}+\frac{\Lambda}{3}$, one can show that Einstein's static model is {unstable}, that is, {with a simple sneeze his Universe collapses}.  This was the real problem of that model, not actually the fact that it did not describe the expansion of the Universe. The case is that Einstein did not realize this problem, that was later noted by Lemaître and by Eddington, among others.

\

To further show the instability, we combine the acceleration and Friedmann equations, to obtain
\begin{eqnarray}
\ddot{a}=-\frac{\dot{a}^2}{2a}-\frac{1}{2a}+\frac{\Lambda a}{2},
\end{eqnarray}
which, as a dynamical system, can be written  as
\begin{eqnarray}
\left\{ \begin{array}{ccc}
  \dot{a} &=  & b \\
  \dot{b} &= &-\frac{b^2}{2a}-\frac{1}{2a}+\frac{\Lambda a}{2},
\end{array}\right.
\end{eqnarray}
where $b=\dot{a}$. It is clear that the Einstein solution corresponds to the fixed point $a=\Lambda^{-1/2}$
and $b=0$.

\

To study the stability of this fixed point, one can linearize the system around it, thus obtaining the matrix
\begin{eqnarray}
\left(\begin{array}{cc}
  0   & 1 \\
 \Lambda    &  0
\end{array}
\right),
\end{eqnarray}
which has a negative determinant, equal to $-\Lambda$; what means that the fixed point is a saddle point, and thus, unstable.

\

Presently, however, the introduction of the CC by Einstein is no longer seen as a horrible mistake but, quite on the contrary, to have had extremely  positive consequences. Namely, now that we know the universe expansion accelerates, the CC could be a most natural candidate for dark energy, to explain the current cosmic acceleration within the standard cosmological model, without any extra addition. To this end,
we consider a flat Universe filled by matter and with the CC.  Since $\rho$ scales as $a^{-3}$ and $\rho_{\Lambda}$ is constant, the constituent equations have the fixed point $\rho=0$, $H=\sqrt{\frac{\Lambda}{3}}$.
This is {de Sitter's} solution, which naturally appears at late times, and since $\dot{H}=0$, i.e., $w_{eff}=-1$, this means that it truly depicts an accelerating Universe.

\

At present, $70\%$ of the energy  density of the Universe is dark and the ordinary matter/energy amount represents some $30\%$ of the total, only. Now, by using the CC  we  have $H_0^2\cong \frac{\Lambda}{3}$. And,  since  $H_0\sim 6\times 10^{-61} M_{pl}$, we obtain a {very small value for the
CC, namely $\Lambda\sim 10^{-120} M_{pl}^2.$} Involving quantum considerations, this number appears to be extremely small, when we compare it with the expected contributions to the CC coming from the unavoidable quantum vacuum fluctuations of the different fields present in the Universe. In order to describe our present Universe, using the CC as a source of dark energy,  we have to fine-tune the CC extremely well, down some hundred orders of magnitude (what has been sometimes called the highest discrepancy between theory and observations ever encountered in physics).

\subsection{Quintessential Inflation}

The main idea in quintessential inflation goes as follows.
{The {inflaton field} could actually be  responsible not only for the  {very early, but also for the late-time acceleration} of the universe. To obtain a successful reheating stage, 
an {abrupt phase transition} must occur from the end of {inflation} to the beginning of {kination} (the epoch when  the energy density of the field was (almost) exclusively kinetic, i.e., $w_{eff}=1$). There, the adiabatic evolution is broken in order to create enough {superheavy particles}, whose energy density ($\langle\rho\rangle\sim a^{-3}$) will eventually dominate the one of the inflaton field ($\rho_{\varphi}\sim a^{-6}$) after {decaying into lighter particles},  in order to match with the hot BB conditions and conveniently enter into the radiation phase in a smooth way (the kination phase ends or the radiation era starts when $\rho_{\varphi}\sim\langle\rho\rangle$). Then, the universe slowly cools down and particles become non-relativistic, thus entering in the matter domination era.

Finally, "close" to present time, the remaining energy density of the inflaton field starts to dominate once again, as a new form of dark energy termed {\it quintessence}, which is able to reproduce {the current cosmic acceleration}, dominating again the energy balance, but now in a much more equilibrated way. The questions: why this is so? and, why does this happen precisely now? are very important ones, and the present standard cosmological model has been unable to answer them up to now.

\begin{remark}
The unification of the early an late time acceleration of our universe could also be obtained in others theories such as modified gravity \cite{nojiri}, in $F(R)$ gravity \cite{nojiri1} or in $F(R,T)$ gravity \cite{nojiri2}, where here $T$ denotes the trace of the stress tensor.
\end{remark}

{An important observation } is also that, owing to the kination regime, the number of "last" e-folds is larger than in the case of standard inflation: in most of the models it ranges, more or less,  between 60 and 70. Effectively, in\cite{review}
 for a model of QI the number of e-folds is given by 
\begin{eqnarray}
 N+\ln  N\cong 55
-\frac{1}{3}\ln\left(\frac{T_{reh}}{M_{pl} } \right),
\end{eqnarray}
and
taking into account that the scale of  nucleosynthesis is $1$ MeV \cite{gkr} and in order to avoid the late-time decay of gravitational relic products such as moduli fields or gravitinos  which could jeopardise  the nucleosynthesis success, one needs temperatures lower than $10^9$ GeV \cite{lindley, eln}. So, we will assume that $1 \mbox{ MeV}\leq T_{reh}\leq 10^9 \mbox{ GeV}$, which leads to constrain the number of e-folds to $58\lesssim N\lesssim 67$.

}

\

The first model of Quintessential Inflation (QI),  was introduced by Peebles and Vilenkin in their seminal paper entitled {\it Quintessential Inflation}  \cite{pv} (see \cite{hap18} for a review and \cite{A,D,E,F,G,HH,II,JJ,bettoni,dimo} for other Quintessential Inflation models
such as exponential models as in \cite{G}, where, to match with the current observational data, the authors assume that the inflaton field is non-minimally coupled with massive neutrinos), and it is defined by the potential:
\begin{eqnarray}\label{PV}
V(\varphi)=\left\{\begin{array}{ccc}
\lambda \left(\varphi^4+  M^4 \right)& \mbox{for} & \varphi\leq 0\\
\lambda\frac{M^8}{\varphi^{4}+M^4} &\mbox{for} & \varphi\geq 0.\end{array}
\right.
\end{eqnarray}
Here, $\lambda$  is a dimensionless parameter and $M\ll M_{pl}$ is a very small mass, as compared with the reduced Planck mass. An abrupt phase transition takes place at $\varphi=0$, 
where the fourth derivative of $V$ is discontinuous.

\

The first part of the potential, {the quartic potential}, is the one responsible for inflation, while the quintessence tail, {the inverse power law potential} is responsible for the current cosmic acceleration.

\

As we will see,  {$\lambda \cong 9\times 10^{-11}$} is obtained from the power spectrum of scalar perturbations and 
{$M\sim 200$ TeV} has to be calculated numerically, using the observational data $\Omega_{\varphi,0}\equiv \frac{\rho_{\varphi,0}}{3H_0^2M_{pl}^2}\cong 0.7$.

 \

 It is here important to recall that the following quantities, which we have already defined, can be actually measured:
  \begin{enumerate}
     \item  The power spectrum of scalar perturbations
     \begin{eqnarray}
     P_{\zeta}=\frac{H_*^2}{8\pi^2 M_{pl}^2\epsilon_*}\sim 2\times 10^{-9},
     \end{eqnarray}
     where the star means that the quantities are evaluated at the horizon crossing.
     \item The spectral index, $n_s\cong 1-6\epsilon_*+2\eta_*$. Its central value is $n_s\cong 0.9649$.
     \item The ratio of tensor to scalar perturbations, $r=16\epsilon_*$. Observational data lead to the constraint $r\leq 0.1$.
 \end{enumerate}

 \

As an example, for a quartic potential $V(\phi)=\lambda \varphi^4$,  the inflationary piece of the PV model, leads to the relation
\begin{eqnarray}n_s=1-\frac{3}{16}r.\end{eqnarray}
However, recent observational data yield
$n_s=0.9649\pm 0.0042$, at $1\sigma$ confidence level, and $r\leq 0.1$. This means that the Peebles-Vilenkin model is not compatible with the current observational data at $2\sigma$ C.L., because from the model one gets the bound: $r\geq 0.1424$.
 
\subsection{Improved versions of QI}

{ {An improved version of QI is {Higgs} Inflation $+$ Inverse Power Law}}, which is given by the potential
\begin{eqnarray}\label{PVimproved}
V(\varphi)=\left\{\begin{array}{ccc}
\lambda M_{pl}^4\left(1-e^{\sqrt{\frac{2}{3}}\frac{\varphi}{M_{pl}}}\right)^2 + \lambda M^4 & \mbox{for} & \varphi\leq 0\\
\lambda\frac{M^8}{\varphi^{4}+M^4} &\mbox{for} & \varphi\geq 0.\end{array}
\right.
\end{eqnarray}

\

For this potential, it is easy to  show that  one gets the relation
\begin{eqnarray}r=3(1-n_s)^2,\end{eqnarray}
which  perfectly fits in the joint contour at $2\sigma$ C.L., because in that case we have $r\leq 0.006$.

 \
 
Now, we  calculate the values of the two parameters of the model.
Note first that the scale of inflation, i.e., the value of the Hubble parameter at the horizon crossing is  $H_{*}^2\cong \frac{\lambda}{3}M_{pl}^2$. Then, the power spectrum for scalar perturbations ${\mathcal P}_{\zeta}=\frac{H_{*}^2}{8\pi^2 M_{pl}^2\epsilon_*}\sim 2\times 10^{-9}$, 
 the formula $r=3(1-n_s)^2$,  the relation $r=16\epsilon_*$ and the central value of the spectral index, $n_s\cong 0.96$, one can show that ${\lambda\cong 9\times 10^{-11}}$.

\

Finally,  from the observational data  $\Omega_{\varphi,0}=\frac{V(\varphi_0)}{3H_0^2 M_{pl}^2}\cong 0.7$, with $\varphi_0\cong 30 M_{pl}$ (we will see later that this is approximately the value of the scalar field at present time), one gets that 
{ $M\sim 10^5$ GeV}.

 \
 
It is important to remark that both the {Peebles-Vilenkin } model, and also this improved version,  should be just viewed as  phenomenological models, as first steps to understand QI.  
More physically grounded models are the following.

\subsubsection{Lorentzian Quintessential Inflation.-} 
Based on the  Lorentzian (or {Cauchy}, for mathematicians) distribution,  one considers the following  ansatz 
\begin{eqnarray}\label{ansatz}
\epsilon(N)=\frac{\xi}{\pi}\frac{\Gamma/2}{N^2+\Gamma^2/4},
\end{eqnarray}
where $\epsilon$ is the  slow-roll parameter,  $N$  the number of e-folds, and $\xi$ and $\Gamma$ the amplitude and width of the Lorentzian distribution,  respectively.

\

From the previous ansatz, one gets the  potential
\cite{benisty,benisty1,benisty2}
\begin{eqnarray}\label{LQI}
V(\varphi)=\lambda M_{pl}^4\exp\left[-\frac{2\gamma}{\pi}\arctan\left(\sinh\left(\gamma\varphi/M_{pl} \right)  \right)\right],
\end{eqnarray}
 where $\lambda$ is a dimensionless parameter and  $\gamma$ is defined by
$$\gamma\equiv \sqrt{\frac{\pi}{\Gamma \xi}}.$$ 
The model depends on these two parameters, and in order to match it with current observational data, one has to impose them to take the values $\lambda\sim 10^{-69}$ and $\gamma\cong 122$. This leads then to a successful inflation model that, at late times, yields an eternal acceleration  with effective EoS parameter equal to $-1$. It is thus indistinguishable from the simple CC in this regime.

\subsubsection{$\alpha$-attractors in Quintessential Inflation.-}  

In that case, the corresponding potential, combined with  a standard   exponential potential, is obtained from a Lagrangian motivated by super-gravity and corresponding to a non-trivial 
K\"ahler manifold.

 \

\begin{figure}[ht]
\begin{center}
\includegraphics[scale=0.7]{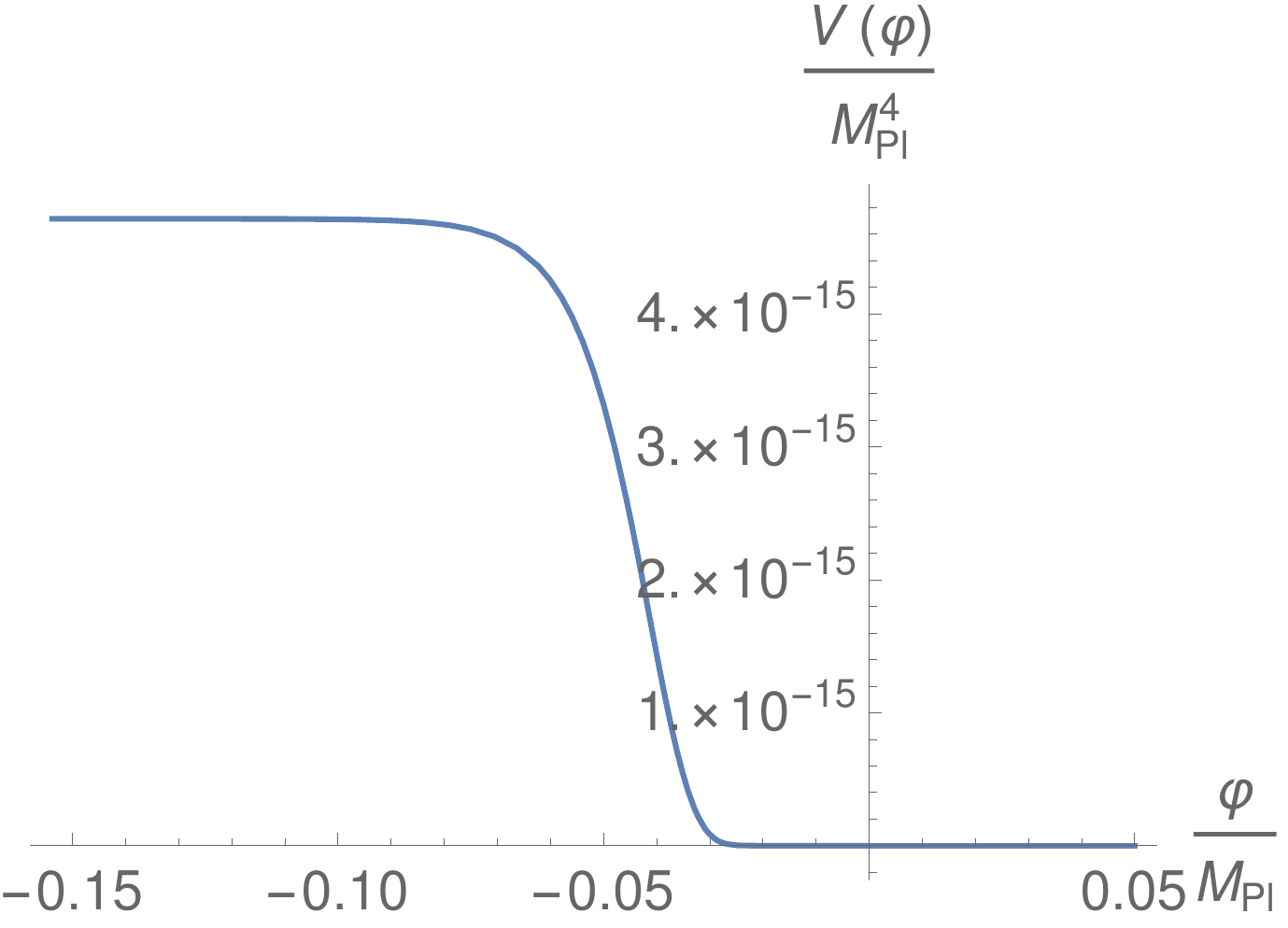}
\end{center}
\caption{  The $\alpha$-attractor potential. In Lorentzian QI the corresponding potential has a similar shape.
}
\label{fig:pot}
\end{figure}

The Lagrangian provided by super-gravity theories is \cite{vardayan}
\begin{eqnarray}\label{lagrangiansup}
\mathcal{L}=\frac{1}{2}\frac{\dot{\phi}^2}{(1-\frac{\phi^2}{6\alpha}  )^2}M_{pl}^2-\lambda M_{pl}^4 e^{-\kappa \phi},
\end{eqnarray}
where $\phi$ is a dimensionless scalar field, and $\kappa$ and $\lambda$ are positive dimensionless constants.
In order that the kinetic term acquires the canonical  form, i.e., 
$\frac{1}{2}{\dot{\varphi}^2}$, one can redefine the scalar field as follows,
\begin{eqnarray}
\phi= \sqrt{6\alpha}\tanh\left(\frac{\varphi}{\sqrt{6\alpha}M_{pl}}  \right),
\end{eqnarray}
thus obtaining the following potential \cite{K,benisty3},
\begin{eqnarray}\label{alpha}
V(\varphi)=\lambda M_{pl}^4e^{-n\tanh\left(\frac{\varphi}{\sqrt{6\alpha}M_{pl}} \right)},
\end{eqnarray}
depicted in Figure \ref{fig:pot}, where we have introduced the notation $n \equiv\kappa \sqrt{6\alpha}$; and  by taking  $\alpha\sim 10^{-2}$,  chosen  to match with observational data, we obtain the current cosmic acceleration when 
$n\sim 10^2$ and $\lambda \sim 10^{-66}$.

\

Finally, for $\alpha$-attractors the spectral index and the tensor/scalar ratio, as a function of the number
of e-folds, have the following simple form
\begin{eqnarray}\label{power}
{ n_s\cong 
1-\frac{2}{{N}}, \qquad  r\cong\frac{12\alpha}{{ N}^2}},
\end{eqnarray}
which, for the usual number of e-folds in Quintessential Inflation ($65\leq N\leq 75$), matches correctly with the observational data $n_s=0.9649\pm0.0042$ and $r\leq 0.1$ at $1$-$\sigma$ confidence level.

Effectively, For $N=65$ the value of the ratio of tensor to scalar perturbations is very small around $r\cong 10^{-3}\alpha$, and smaller for $N>65.$

\subsection{Evolution of the dynamical system}

To understand the evolution of the Universe in QI, we deal with  the model (\ref{PVimproved}). First, we calculate the energy density at the end of inflation. Since inflation ends when 
$\epsilon\cong -\frac{\dot{H}}{H^2}=1\Longrightarrow w_{eff}=-1/3$, one has $\rho_{END}=\frac{3}{2}V(\varphi_{END})$. 

\

On the other hand, since $\epsilon=\frac{M_{pl}^2}{2}\left(\frac{V_{\varphi}}{V} \right)^2,$  at the end of inflation  one has
$\varphi_{END}=\frac{1}{2}
\sqrt{\frac{3}{2}}
\ln\left( 3/4 \right) M_{pl}.$

\

 Note that the  kination regime starts  approximately for $\varphi_{kin}=0>\varphi_{END}$ when the potential is negligible. Assuming, as usual, that {there is no energy drop} between the end of inflation and the beginning of kination, we get $\dot{\varphi}_{kin}=\sqrt{3V(\varphi_{END})}$. Then, the initial conditions at the beginning of kination are
\begin{eqnarray}\left(\varphi_{kin}=0, \dot{\varphi}_{kin}=\sqrt{\frac{3\lambda}{4}}(2-\sqrt{3})M_{pl}^2\right) \Longrightarrow H_{kin}={\frac{\sqrt{\lambda}}{2\sqrt{2}}}(2-\sqrt{3})M_{pl},\end{eqnarray}
and thus, {$H_{kin}\cong 10^{-6} M_{pl}$}.

\

During kination, the effective EoS parameter is given by $w_{eff}=1$, because all energy density is kinetic. So, during this period the Hubble rate evolves as $H(t)=\frac{1}{3t}$. Then, since the value of the potential is very small as compared with the kinetic energy, one can safely disregard it in the FE, thus obtaining
\begin{eqnarray}
\frac{\dot{\varphi}^2}{2}=\frac{M_{pl}^2}{3t^2}\Longrightarrow \dot{\varphi}(t)=\sqrt{\frac{2}{3}}\frac{M_{pl}}{t}
\Longrightarrow 
\varphi(t)=\sqrt{\frac{2}{3}}M_{pl}\ln \left( \frac{t}{t_{kin}} \right).\end{eqnarray}
Therefore, we end up with
\begin{eqnarray}\left(\varphi(t)=\sqrt{\frac{2}{3}}M_{pl}\ln \left( \frac{H_{kin}}{H(t)} \right), \dot{\varphi}(t)=\sqrt{6} H(t)M_{pl}\right).\end{eqnarray}

\

Now, for simplicity, {we shall assume that the created particles during the phase transition} from the end of inflation to the beginning of kination {decay in lighter ones before the end of kination occurs, and that they thermalize almost instantaneously}. Under these circumstances, the Universe becomes reheated at the end of the kination phase (the energy density of the inflaton is the same as the one of the created particles), just when the energy density of the created particles start to dominate. As a consequence, at reheating time, we have
\begin{eqnarray}\left(\varphi_{rh}=\sqrt{\frac{2}{3}}M_{pl}\ln \left( \frac{H_{kin}}{H_{rh}} \right), \dot{\varphi}_{rh}=\sqrt{6} H_{rh}M_{pl}\right).\end{eqnarray}
And using that $H_{rh}^2\cong \frac{2\rho_{rh}}{3M_{pl}^2}$, where $\rho_{rh}=\frac{\pi^2}{30}g_*T_{rh}^4$ is the energy density of radiation ({Stefan-Boltzmann's} law), one can write the value of the inflaton field and its derivative as a function of the reheating temperature.

\

Next, during radiation domination $w_{eff}=1/3$, we have $H(t)=\frac{1}{2t}$.
Taking into account that,  in the radiation epoch, the potential energy may be disregarded too (since it is negligible as compared with the kinetic one),  the CE becomes
\begin{eqnarray}
\ddot{\varphi}+\frac{1}{2t}\dot{\varphi}=0.
\end{eqnarray}
Integrating now this equation, one can see that, during the radiation domination epoch, one has
\begin{eqnarray*}\varphi(t)=\varphi_{rh}+2\dot{\varphi}_{rh}t_{rh}\left(1-\sqrt{\frac{t_{rh}}{t}}\right)=
\varphi_{rh}+\frac{\dot{\varphi}_{rh}}{H_{rh}}
\left(1-\sqrt{\frac{H(t)}{H_{rh}}}\right) \quad \mbox{ and } \quad
\dot{\varphi}(t)=\dot{\varphi}_{rh}\left( \frac{H(t)}{H_{rh}} \right)^{3/2}.
\end{eqnarray*}
Therefore, at the matter-radiation equality we will have the following initial conditions
$$
\left(\varphi_{eq}=\varphi_{rh}+\frac{\dot{\varphi}_{rh}}{H_{rh}} 
\left(1-\sqrt{\frac{H_{eq}}{H_{rh}}}\right), \dot{\varphi}_{eq}=\dot{\varphi}_{rh}\left( \frac{H_{eq}}{H_{rh}} \right)^{3/2}\right).
$$

\

And at matter-radiation equality, we have 
$H_{eq}^2\cong \frac{2\rho_{eq}}{3M_{pl}^2}$, where $\rho_{eq}=\frac{\pi^2}{30}g_{eq}T_{eq}^4$, 
being the number of degrees of freedom: $g_{eq}=3.36$.

\

Now, it is important to know the corresponding initial conditions. From the central values, as obtained by {\it the Planck collaboration},   for  the {cosmological red-shift} at matter-radiation equality  {$z_{eq}\equiv \frac{a_0}{a_{eq}}-1=3365$}, and from the present value of the ratio of the matter energy density to the critical one {$\Omega_{m,0}\equiv \frac{\rho_{m,0}}{3H_0^2M_{pl}^2}=0.308$},
one can deduce that  the present value of the matter energy density is {$\rho_{m,0}=3H_0^2M_{pl}^2\Omega_{m,0}=3.26\times 10^{-121} M_{pl}^4$}, and at the matter-radiation equality one should have {$\rho_{m,eq}=\rho_{r,eq}=\rho_{m,0}(1+z_{eq})^3=
4.4\times 10^{-1} \mbox{eV}^4$}, and thus from the Stefan-Boltzmann law,  $T_{eq}\sim 3\times 10^{-28} M_{pl}$.

\

Since  $H_0\sim 6\times 10^{-61} M_{pl}$, choosing a viable temperature, as for example $T_{rh}= 10^9$ GeV, one has
\begin{eqnarray}\varphi_{eq}=\varphi_{rh}+\sqrt{6}M_{pl} \quad \mbox{and} \quad \frac{\dot{\varphi}_{eq}}{H_0M_{pl}}\cong 0.\end{eqnarray}
That is,
\begin{eqnarray}\varphi_{eq}=\sqrt{\frac{2}{3}}M_{pl}\ln\left( \frac{3\sqrt{5\lambda}(2-\sqrt{3})M_{pl}^2}{2\pi T_{rh}^2}   \right)+\sqrt{6}M_{pl},\end{eqnarray}
and inserting the values of $\lambda$ and $T_{rh}$, one finally gets
\begin{eqnarray}\label{initialconditions1}\varphi_{eq}\cong 27.29 M_{pl} \quad \mbox{and} \quad \frac{\dot{\varphi}_{eq}}{H_0M_{pl}}\cong 0.\end{eqnarray}

\subsubsection{The dynamical system}

In order to obtain the dynamical system for this model, we 
introduce the following dimensionless variables
 \begin{eqnarray}\label{variables}
 x=\frac{\varphi}{M_{pl}}, \qquad y=\frac{\dot{\varphi}}{H_0 M_{pl}}.
 \end{eqnarray}
  Using the variable $N$ as time,
 $N = \ln \left( \frac{a}{a_0} \right)$, and from the CE $\ddot{\varphi}+3H\dot{\varphi}+V_{\varphi}=0$, one can build  the following  non-autonomous dynamical system:
 \begin{eqnarray}\label{system}
 \left\{ \begin{array}{ccc}
 x^\prime & =& \frac{y}{\bar H}~,\\
 y^\prime &=& -3y-\frac{\bar{V}_x}{ \bar{H}}~,\end{array}\right.
 \end{eqnarray}
 where the prime means derivative with respect to $N$, $\bar{H}=\frac{H}{H_0}$   and $\bar{V}=\frac{V}{H_0^2M_{pl}^2}$.

\

Moreover, the FE now reads  
 \begin{eqnarray}\label{friedmannquintessence}
 \bar{H}(N)=\frac{1}{\sqrt{3}}\sqrt{ \frac{y^2}{2}+\bar{V}(x)+ \bar{\rho}_{rad}(N)+\bar{\rho}_{matt}(N) }~,
 \end{eqnarray}
where we have introduced the following dimensionless energy densities
 $\bar{\rho}_{r}=\frac{\rho_{rad}}{H_0^2M_{pl}^2}$ and 
 $\bar{\rho}_{m}=\frac{\rho_{matt}}{H_0^2M_{pl}^2}$, being 
\begin{eqnarray}
\rho_{matt}(N)=\rho_{matt,eq}e^{3(N_{eq}-N)} \quad \mbox{and} \quad \rho_{rad}(N)=\rho_{rad,eq}e^{4(N_{eq}-N)}, \end{eqnarray}
the corresponding matter and radiation energy densities.

\

\begin{figure}[ht]
\begin{center}
\includegraphics[scale=0.5]{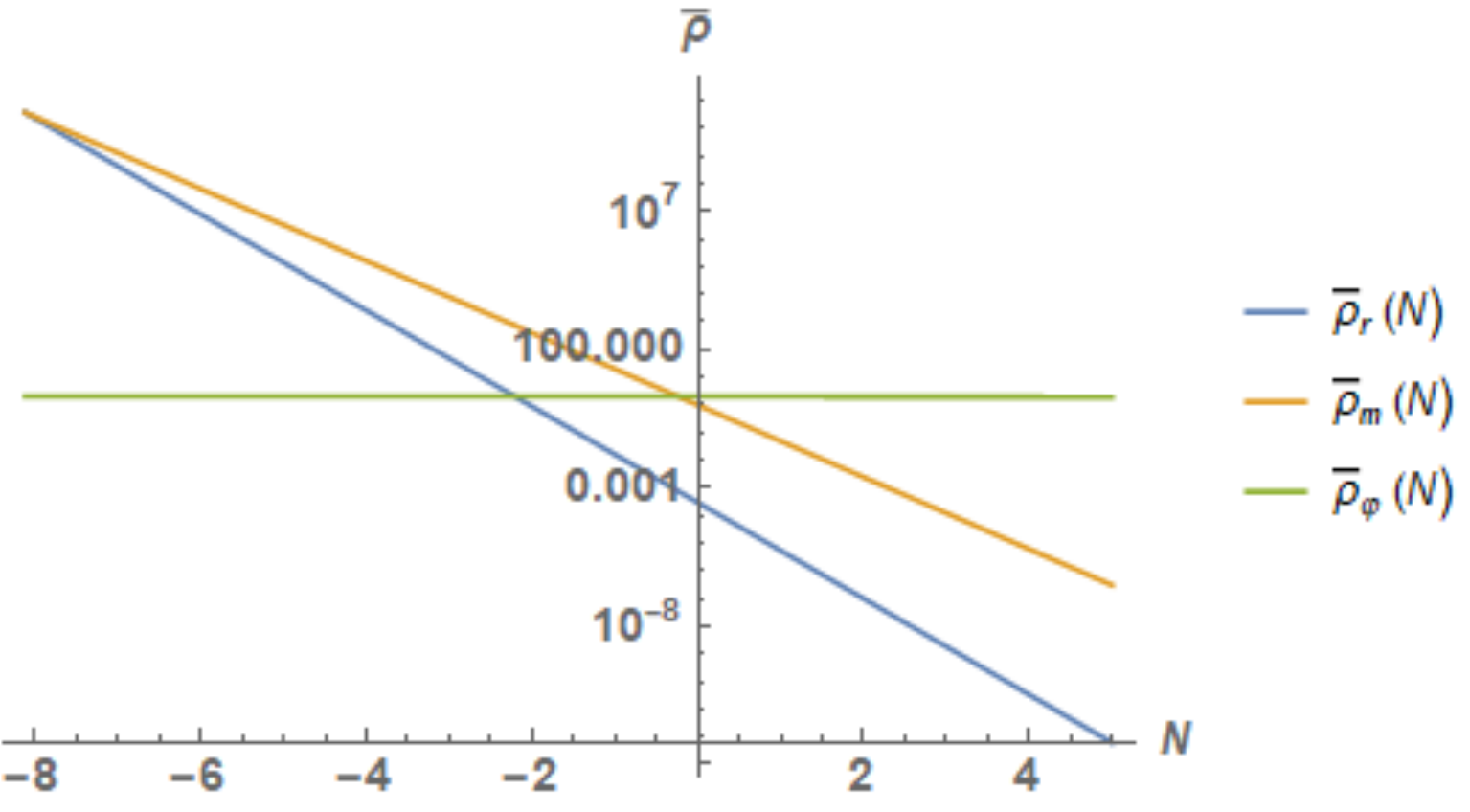}
\includegraphics[scale=0.5]{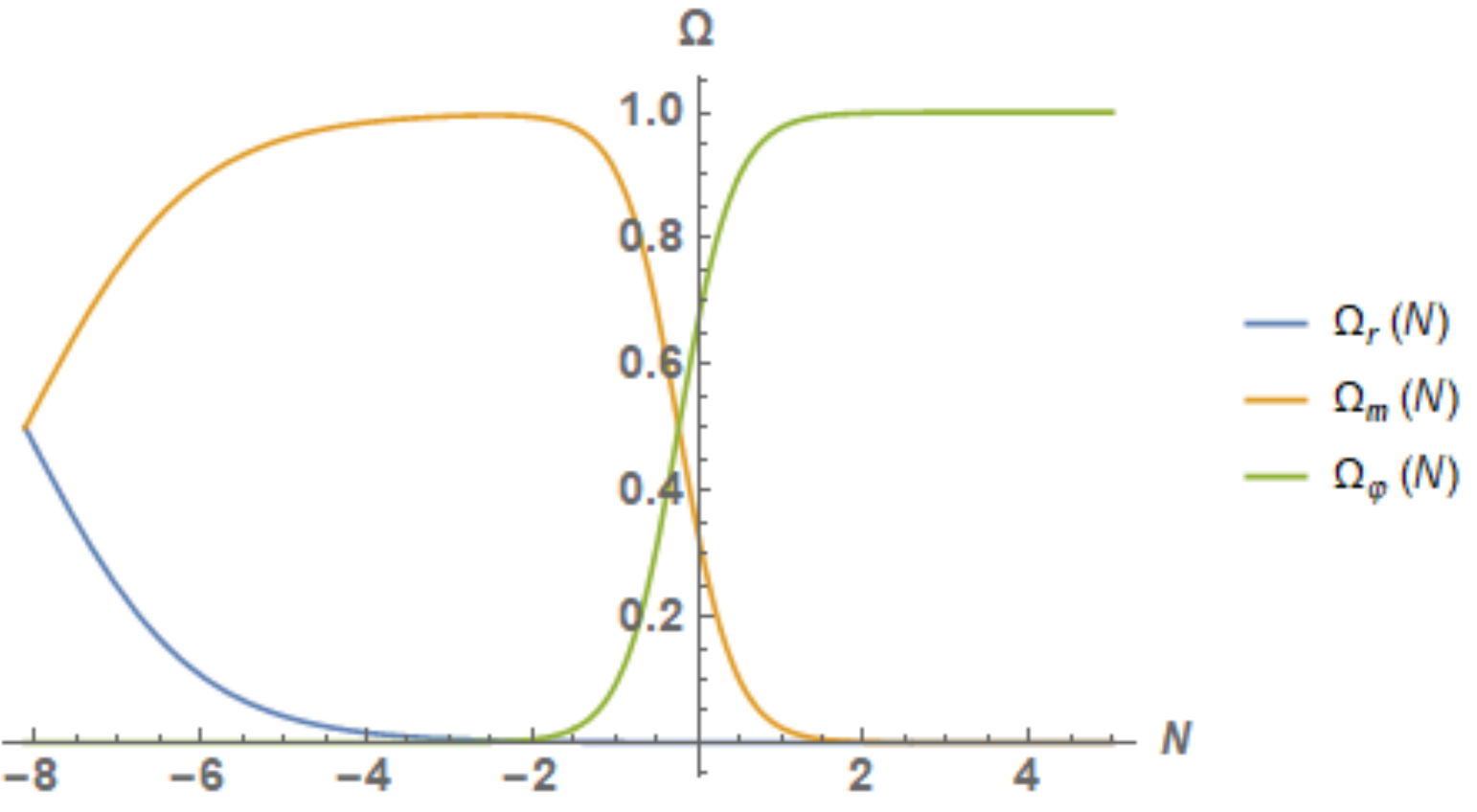}
\end{center}
\caption{ Evolution of $\{\bar{\rho}_B(N)\}_{B=r,m,\varphi}$ and
$\{\Omega_B(N)\}_{B=r,m,\varphi}$. 
}
\label{fig:densitats}
\end{figure}

Integrating numerically the dynamical system with initial conditions $x_{eq}=27$ and $y_{eq}=0$
obtained in (\ref{initialconditions1}) and imposing the condition $\bar{H}(0)=1$ which only holds for $M\cong 10^5$ GeV (the value we have previouly obtained), we get the result obtained in Figure \ref{fig:densitats}, where $\Omega_B=\frac{H_0^2\bar{\rho}_B}{3H^2}$ being $B=r,m, \varphi$, is the ratio of the energy density to the critical one. We  see that, at the present time $N=0$,  dark energy dominates, and that for this model it will continue dominating for ever.

\

\begin{figure}[ht]
\begin{center}
\includegraphics[scale=0.7]{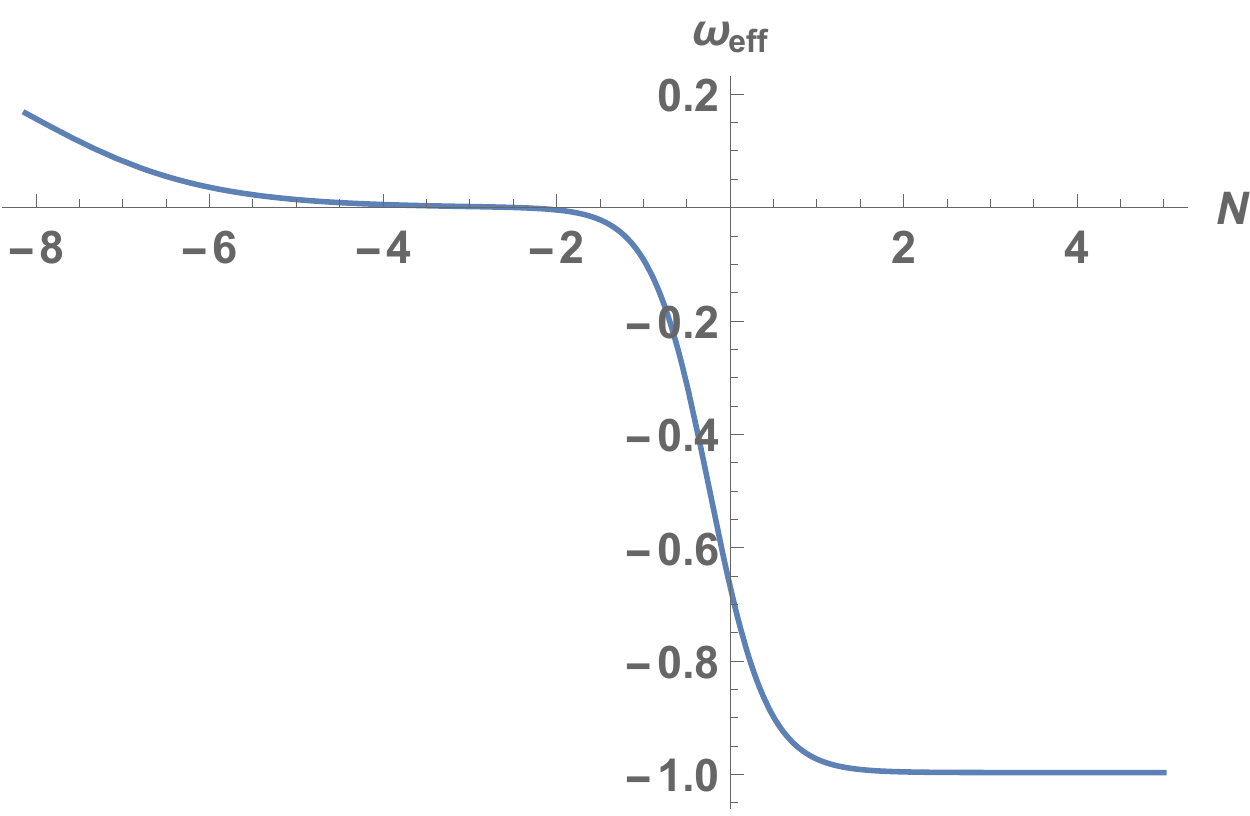}
\end{center}
\caption{ Evolution of 
$w_{eff}$. }
\label{fig:weff}
\end{figure}

Finally from Figure \ref{fig:weff} one can see the evolution of the effective EoS parameter
$w_{eff}=-1-\frac{2\dot{H}}{3H^2}$. Integration starts at the beginning of the matter-radiation equality. We see that,
at present time, $w_{eff}<-1/3$, so {our Universe accelerates}. In fact, at late times $w_{eff}$ evolves towards $-1$, which means that the Universe will always be in accelerated expansion in the future.

\section{The reheating mechanism}
To understand the reheating mechanism via particle production we shall need, first of all, some basic notions of quantum mechanics. To "warm up", the best example to study is the quantum harmonic oscillator.

\subsection{The harmonic oscillator}
We consider the equation of a $1$-dimensional harmonic oscillator
\begin{eqnarray}\label{din}
\ddot{x}+\omega^2x=0,
\end{eqnarray}
whose Hamiltonian is given by
\begin{eqnarray}
{\mathcal H}=\frac{1}{2}(p^2+\omega^2x^2),
\end{eqnarray}
where $p=\dot{x}$ denotes the corresponding momentum.

\

The dynamical equation (\ref{din}) can be written as a Hamiltonian system
\begin{eqnarray}
\left\{ \begin{array}{ccccc}
\dot{x}&=& \partial_p {\mathcal H}&=& \left\{x,{\mathcal H}\right\}\\
\dot{p}&=& -\partial_x {\mathcal H}&=& \left\{p,{\mathcal H}\right\},
\end{array}\right.
\end{eqnarray}
where we have introduced the Poisson bracket $\left\{ A, B\right\}\equiv 
\partial_x A \partial_p B-\partial_p A \partial_x B$.

\

Next, we consider its quantum version. For this, following the correspondence principle, we have to replace  the dynamical  variables by the operators: $x\rightarrow \hat{x}$ and $p\rightarrow \hat{p}$, and the Poisson bracket has to be replaced by a commutator, namely,
\begin{eqnarray}
\left\{ A, B\right\}\rightarrow -i[\hat{A},\hat{B}]\equiv -i(\hat{A}\hat{B}-\hat{B}\hat{A}).
\end{eqnarray}
Thus, taking into account now that the  canonically conjugate variables satisfy $\left\{ x, p\right\}=1$, one obtains its quantum analogue $[\hat{x},\hat{p}]=i$, and 
using the Heisenberg picture, where the operators are time-dependent, the quantum version of the Hamilton equations read 
\begin{eqnarray}
\left\{ \begin{array}{ccc}
\dot{\hat{x}}&=&  -i[\hat{x},\hat{\mathcal H}]\\
\dot{\hat{p}}&=&  -i[\hat{p},\hat{\mathcal H}].
\end{array}\right.
\end{eqnarray}

\

On the other hand, in the Schr\"odinger picture, which is actually equivalent to the Heisenberg one, the operators do not depend on time; thus, time dependence falls off the wave function. In fact, taking the most usual representation,
\begin{eqnarray}
\hat{x}=x,\quad \hat{p}=-i\partial_x, \quad \mbox{and} \quad \hat{E}=i\partial_t,
\end{eqnarray}
the quantum version of the classical equation $E=\mathcal{H}$ (the energy being equal to the Hamiltonian) reads
\begin{eqnarray}
i\partial_t\Phi(t,x)=\frac{1}{2}\left(-\partial_{xx}\Phi(t,x)+\omega^2x^2\Phi(t,x)  \right),
\end{eqnarray}
which is the well-known {\it Schr\"odinger equation} for the harmonic oscillator.

\

The eigenvalues of the quantum Hamiltonian are not difficult to obtain. In fact, the ground state is reached by looking for an eigenfuction of the form $e^{-ax^2}$. Inserting it in the equation $\lambda \Phi=\hat{H}\Phi$, one gets
\begin{eqnarray}
\Phi_0(x)= \frac{1}{(\omega^2\pi)^{1/4}}e^{-\frac{\omega}{2}x^2} \mbox{ with } \lambda_0=\frac{\omega}{2},
\end{eqnarray}
where we have normalized the ground state.

\

To obtain the other eigenvalues, we diagonalize the Hamiltonian, by introducing the so-called, creation and annihilation operators:
\begin{eqnarray}
\hat{a}^{\dagger}=\frac{1}{\sqrt{2\omega}}(\omega\hat{x}-i\hat{p})\quad \mbox{and} \quad 
\hat{a}=\frac{1}{\sqrt{2\omega}}(\omega\hat{x}+i\hat{p}).
\end{eqnarray}
Tanking into account the commutation relation between canonically conjugate variables, it is not difficult to show that the creation and annihilation operators satisfy the relation 
$[\hat{a},\hat{a}^{\dagger}]=1$. And then, the expression of the Hamiltonian as a function of both operators is
\begin{eqnarray}
\hat{\mathcal H}=\frac{\omega}{2}(\hat{a}\hat{a}^{\dagger}+\hat{a}^{\dagger}\hat{a})=
\omega\left(\hat{a}^{\dagger}\hat{a}+\frac{1}{2}  \right),
\end{eqnarray}
where we have used the commutation relations.

\

Finally, with the creation operator $\hat{a}^{\dagger}$ we build  the whole set of eigenstates:
\begin{eqnarray}
\Phi_n=\frac{1}{\sqrt{n!}}(\hat{a}^{\dagger})^n \Phi_0,
\end{eqnarray}
whose corresponding eigenvalues are $\lambda_n=\omega\left(n+\frac{1}{2}\right).$

\

Now, coming back to the Heisenberg picture,  the dynamical equations for the creation an annihilation operators read
\begin{eqnarray}\dot{\hat{a}}=-i[\hat{a}, \hat{\mathcal H}]=-i\hat{a}\quad \mbox{and}\quad \dot{\hat{a}}^{\dagger}=-i[\hat{a}^{\dagger}, \hat{\mathcal H}]=i\hat{a}^{\dagger},\end{eqnarray}
and their solution is 
\begin{eqnarray}\label{sol}
\hat{a}(t)=e^{-i\omega(t-t_i)}\hat{a}(t_i)\quad \mbox{and} \quad \hat{a}^{\dagger}(t)=e^{i\omega(t-t_i)}
\hat{a}^{\dagger}(t_i).
\end{eqnarray}
Thus, one gets
\begin{eqnarray}
\hat{x}(t)=\chi(t)\hat{a}(t_i)+\chi^*(t)\hat{a}^{\dagger}(t_i) \quad \mbox{and}\quad
\hat{p}(t)=\dot{\chi}(t)\hat{a}(t_i)+\dot{\chi}^*(t)\hat{a}^{\dagger}(t_i),
\end{eqnarray}
where $\chi(t)=\frac{1}{\sqrt{2\omega}}e^{-i\omega(t-t_i)}$ is the positive frequency mode.

\

Finally, form the commutation relation $[\hat{x}(t),\hat{p}(t)]=1$, one can check that the Wronskian of  $\chi(t)$ and its conjugate, reads 
\begin{eqnarray}
{\mathcal W}[\chi, \chi^*]\equiv\chi \dot{\chi}^*-\dot{\chi}\chi^*=i,
\end{eqnarray}
which may seem obvious for the mode $\chi(t)=\frac{1}{\sqrt{2\omega}}e^{-i\omega(t-t_i)}$; but when we deal with a time-dependent frequency, this relation will be essential.

\

A final remark is in order: The ground state always satisfies $\hat{a}(t)\Phi_0=0$, because the creation and annihilation operators evolve as (\ref{sol}), i.e., positive and negative frequencies do not mix. However, as we will see in the next Section, for a time dependent frequency, mixing does happen, and it is a basic ingredient that induces particle production.

\subsection{The harmonic oscillator with a time dependent frequency}

As in the constant case, the quantum Hamiltonian is given by 
\begin{eqnarray}
\hat{\mathcal H}=\frac{1}{2}(\hat{p}^2+\omega^2(t)\hat{x}^2),
\end{eqnarray}
and the quantum equation for $\hat{x}$ in the Heisenberg picture is  again
\begin{eqnarray}
\ddot{\hat{x}}+\omega^2(t)\hat{x}=0.
\end{eqnarray}
Next, we write 
\begin{eqnarray}
\hat{x}(t)=\chi(t)\hat{a}(t_i)+\chi^*(t)\hat{a}^{\dagger}(t_i) \quad \mbox{and}\quad
\hat{p}(t)=\dot{\chi}(t)\hat{a}(t_i)+\dot{\chi}^*(t)\hat{a}^{\dagger}(t_i),
\end{eqnarray}
where the positive frequency modes satisfy the Klein-Gordon equation $\ddot{\chi}+\omega^2(t)\chi=0$, and also the Wronskian condition $\chi\dot{\chi}^*-\dot{\chi}\chi^*=i$. Unfortunately, these modes do not have the simple expression $\chi(t)=\frac{1}{\sqrt{2\omega}}e^{-i\omega(t-t_i)}$ as in the case of constant frequency.

\

Anyway, we will assume that at early times the adiabatic condition $\frac{\dot{\omega}}{\omega^2}\ll 1$ is fulfilled, which, as we will see, does always happen during inflation. In this epoch, one can consider a positive frequency mode, of the form
\begin{eqnarray}
\chi(t)=\frac{1}{\sqrt{2\omega(t)}}e^{-i\int_{t_i}^t \omega(s)ds},
\end{eqnarray}
and which  satisfies
\begin{eqnarray}
\dot{\chi}=\left(-i\omega-\frac{\dot{\omega}}{2\omega}\right)\chi\cong -i\omega \chi,
\quad \mbox{and}\quad
\ddot{\chi}=\left(-\frac{i\dot{\omega}}{2}-\omega^2\right)\chi\cong -\omega^2 \chi.
\end{eqnarray}
Thus, during the adiabatic regime the mode $\chi$ is very close to the solution that satisfies the initial conditions 
\begin{eqnarray}\label{initial}
\chi(t_i)=\frac{1}{\sqrt{2\omega(t_i)}}\quad \mbox{and} \quad \dot{\chi}(t_i)=-i\sqrt{\frac{\omega(t_i)}{2}}.
\end{eqnarray}

\

Next, working in the Heisenberg picture,  we consider the groundstate at time $t_i$, i.e., $\Phi_0(t_i)$. which satisfies $\hat{a}(t_i)\Phi_0(t_i)=0$ -e.g., the positive frequency mode satisfying the initial conditions (\ref{initial})- and we calculate the average, at any time, of the energy with respect the ground state:
\begin{eqnarray}\label{average}
\Phi_0^*(t_i)\hat{\mathcal H}(t)\Phi_0(t_i)=\frac{1}{2}\left(|\dot{\chi}|^2+\omega^2(t)|{\chi}|^2\right).
\end{eqnarray}
At time $t_i$, it coincides with its minimum $\omega(t_i)/2$.

\

Finally, we use {\it the diagonalization method} to look  for modes of the form
\begin{eqnarray}
\chi(t)=\alpha(t)\frac{1}{\sqrt{2\omega(t)}}e^{-i\int_{t_i}^t \omega(s)ds}+\beta(t)\frac{1}{\sqrt{2\omega(t)}}e^{i\int_{t_i}^t \omega(s)ds}.
\end{eqnarray}
Here, $\alpha(t)$ and $\beta(t)$ are the so-called time-dependent Bogoliubov coefficients.

\

Imposing that the mode satisfies   the condition
\begin{eqnarray}
\dot{\chi}(t)= -i\omega(t)\left(\alpha(t)\frac{e^{-i\int^{t}_{t_i} \omega(s)ds}}{\sqrt{2\omega(t)}}-
\beta(t)\frac{e^{i\int^{t}_{t_i} \omega(s)ds}}{\sqrt{2\omega(t)}}\right),\end{eqnarray}
due to the Wronskian property $\chi\dot{\chi}^*-\dot{\chi}\chi^*=i$, they have to satisfy $|\alpha(t)|^2-|\beta(t)|^2=1$. In addition, in order that the mode is a solution of  the Klein-Gordon equation, these coefficients need satisfy the system of equations
\begin{eqnarray}\label{Bogoliubov}
\left\{ \begin{array}{ccc}
\dot{\alpha}(t) &=& \frac{\dot{\omega}(t)}{2\omega(t)}e^{2i\int^{t}_{t_i} \omega(s)ds}\beta(t)\\
\dot{\beta}(t) &=& \frac{\dot{\omega}(t)}{2\omega(t)}e^{-2i\int^{t}_{t_i}\omega(s)ds}\alpha(t).\end{array}\right.
\end{eqnarray}

\

And inserting this expression into (\ref{average}), one obtains the following "vacuum average"
\begin{eqnarray}
\Phi_0^*(t_i)\hat{\mathcal H}(t)\Phi_0(t_i)=\frac{1}{2}\left(|\dot{\chi}|^2+\omega^2(t)|{\chi}|^2\right)=\omega(t)|\beta(t)|^2+\frac{\omega(t)}{2}.
\end{eqnarray}
This is an example of the well-know diagonalization method, which is used in cosmology to calculate the average energy density of the produced particles.

\

Generally, one subtracts  the term  $\omega(t)/2$,  which corresponds to the minimum energy, thus obtaining 
\begin{eqnarray} \langle \hat{\mathcal H}(t) \rangle\equiv
\Phi_0^*(t_i)\hat{\mathcal H}(t)\Phi_0(t_i)-\frac{\omega(t)}{2} =\omega(t)|\beta(t)|^2.
\end{eqnarray}

\subsection{Gravitational particle production of a massive quantum field conformally coupled to gravity}

Here we consider a massive  scalar field $\phi$ conformally coupled to gravity. Its Lagrangian density is given by \cite{gmmbook}
\begin{eqnarray}
{\mathcal L}=\frac{1}{2}(\partial_{\mu}\phi\partial^{\mu}\phi-m_{\chi}^2\phi^2-\frac{R}{6}\phi^2),
\end{eqnarray}
where $m_{\chi}$ denotes the mass and 
whose corresponding Klein-Gordon equation, obtained form the Euler-Lagrange equation,  reads
\begin{eqnarray}
\left(-\nabla^{\mu}\nabla_{\mu}+m_{\chi}^2+\frac{R}{6}\right)\phi=0
\end{eqnarray}
and, after the change of variables $\phi=\chi/a$ ($a$ denotes, once again, the scalar factor), it acquires the more usual form
\begin{eqnarray}
\chi''-\Delta\chi+m_{\chi}^2a^2\chi=0,
\end{eqnarray}
where the tilde  denotes  derivative with respect to  conformal time, $d\tau=\frac{dt}{a(t)}$.

\

We now work in Fourier space, where the quantum version reads 
\begin{eqnarray}
\hat{\chi}(\tau, {\bf x})=\frac{1}{(2\pi)^{3/2}}\int \left(\hat{a}_{\bf k}(t_i) e^{i {\bf k}.{\bf x}}\chi_k(\tau)+\hat{a}^{\dagger}_{\bf k}(t_i) e^{-i {\bf k}.{\bf x}}\chi_k^*(\tau)\right) d^3{\bf k},
\end{eqnarray}
and the modes $\chi_k$ satisfy the equation of an harmonic oscillator 
\begin{eqnarray}
\chi_k''+\omega_k^2(\tau)\chi_k=0,
\end{eqnarray}
with time-dependent frequency given by  $\omega_k(\tau)=\sqrt{k^2+a^2(\tau)m_{\chi}^2}$.

\

The corresponding average energy density is  \cite{Bunch}
\begin{eqnarray}
\langle\rho(\tau)\rangle=\frac{1}{4\pi^2a^2(\tau)}\int k^2(|\chi_k'|^2+\omega_k^2(\tau)|\chi_k|^2-\omega_k(\tau)   )dk,
\end{eqnarray}
where, as for a single oscillator, we have subtracted the minimum vacuum energy density, i.e., the so-called zero-point oscillations of the quantum vacuum: $\frac{1}{(2\pi)^3a^4(\tau)}\int d^3k  \frac{1}{2} \omega_k(\tau)$.

\

At this point,  it is useful once more to apply the diagonalization method, to get the simple form for the energy density:
\begin{eqnarray}\label{vacuum-energy1}
\langle\rho(\tau)\rangle= \frac{1}{2\pi^2a^4(\tau)}\int_0^{\infty} k^2\omega_k(\tau)|\beta_k(\tau)|^2 dk,
\end{eqnarray}
where the time-dependent Bogoliubov coefficients, $\alpha_k$ and $\beta_k$, satisfy the system (\ref{Bogoliubov}) for each value of $k$.

\

It is important to notice that $|\beta_k(\tau)|^2$ encodes the vacuum polarization effects (the creation and annihilation of pairs of opposite charge) and also the production of particles, which only happens however when the adiabatic evolution breaks down.  
Actually, in quintessential inflation the adiabatic regime is broken during the abrupt phase transition between the end of inflation and the beginning of the kination stage and, just at the very beginning of kination, the adiabatic regime is recovered.  So, the polarization effects disappear and  the $\beta$-Bogoliubov coefficients stabilize at a constant value, which only encodes the production of real particles.

\

Now, the way to calculate (numerically) $|\beta_k(\tau)|^2$ goes as follows:
First of all, we need to integrate numerically the conservation equation for the inflaton field, namely
\begin{eqnarray}\label{conservation3}
\ddot{\varphi}+3H\dot{\varphi}+V_{\varphi}=0,
\end{eqnarray}
where $H=\frac{1}{\sqrt{3}M_{pl}}\sqrt{\frac{\dot{\varphi}^2}{2}+V(\varphi)  }$,
 with initial conditions at some moment during the slow-roll regime. Recall that, at that instant,
the system is in the slow-roll phase and, since this regime is an attractor, one only needs to take initial conditions in the basin of attraction of the slow-roll solution. So, we take initial conditions at the at  horizon crossing,
 i.e., when the pivot scales leaves the Hubble radius
$\varphi_*$ and $\dot{\varphi}_*=-\frac{V_{\varphi}(\varphi_*)}{3H_*}$.

\

Once we  have obtained the evolution of the background, and, in particular, the evolution of the Hubble rate, we compute the evolution of the scale factor. This is given by 
\begin{eqnarray}
a(t)=e^{\int_{t_*}^t H(s)ds},
\end{eqnarray}
where we have chosen, at the horizon crossing,  $a_*=1$.
Once we have the evolution of this scale factor, we just need to numerically integrate the system (\ref{Bogoliubov}).

\

A final remark is in order. When particles are produced, the Friedmann equation gets modified, as follows 
\begin{eqnarray}
H^2=\frac{1}{3M_{pl}^2}(\rho_{\varphi}+\langle\rho\rangle),
\end{eqnarray}
where $\rho_{\varphi}$ denotes the energy density of the inflaton field and $\langle \rho\rangle$ the one of the produced particles.

\subsection{Reheating in the case of $\alpha$-attractors}
Dealing with $\alpha$-attractors (see the potential (31)) one can show that superheavy particles,
 with masses around $m_{\chi}\sim 10^{15}$ GeV, are gravitationally produced during the abrupt phase transition from the end of inflation to the beginning of kination. Note that such heavy masses are needed in order that the polarization effects do not affect the dynamics of the inflaton field during inflation, because it is usually assumed that  inflation starts at GUT scales with $H_{GUT}\sim 10^{14}$ GeV, and it is well know that polarization effects can be neglected when $H\ll m_{\chi}$. 
 
 \
 
 Therefore, the produced $\chi$-particles have energy densities of the order (see for instance \cite{ah2021}):	
{ \begin{eqnarray}
 \langle\rho(\tau)\rangle \sim \frac{m_{\chi}^4}{6\pi^2}e^{-\kappa \alpha m_{\chi}/H_*}
\left(\frac{a_{kin}}{a(\tau)}\right)^3,\end{eqnarray}
}
where $\kappa$ is a dimensionless number of order $1$
(more precisely, for $\alpha=10^{-3}$ numerically one  obtains $\kappa\cong 12.13$; for
$\alpha=10^{-2}$  the calculation yields $\kappa\cong 3.45$;  for $\alpha=10^{-1}$ one gets $\kappa\cong 1.1$; and, finally, for $\alpha=1$ one obtains $\kappa\cong 0.3$), and $H_*\sim 10^{13}$ GeV  is the escale of inflation, i.e., the value of the Hubble parameter at  horizon crossing.

\

What is actually important is that these superheavy particles all decay into lighter ones, to form a relativistic plasma which will eventually dominate the evolution of the universe and  thus, conveniently match the hot Big Bang model. Then, {two different situations may arise}
{\begin{enumerate}
\item The decay takes place before the end of the kination phase (recall that kination ends when $\rho_{\varphi}\sim \langle\rho\rangle$).
	\item  The decay occurs after the end of kination.
\end{enumerate}}

\

Let $\Gamma$ be the decay rate; then the decay is finished when $\Gamma\sim H$, because $H\sim 1/t$, the number of $\chi$-particles decay as $e^{-\Gamma t}$ and the decay is practically  finished when $e^{-\Gamma t}\cong 1/2$.

\

\subsubsection{Decay happens before the end of kination}

In this case we have the two constraints:
\begin{enumerate}
\item  The decay occurs after the beginning of kination, i.e., $\Gamma\leq H_{kin}\sim 6\times 10^{-7} M_{pl}$. This value has been obtained numerically in the case of $\alpha$-attractors, and it agrees with the fact that kination starts immediately after the end of inflation, which in the majority of models ends at the scale $H_{END}\sim 10^{-7}$ GeV.
\item The decay precedes the end of kination, i.e., $\langle\rho_{ dec}\rangle\leq \rho_{\varphi, dec}$.
Taking into account that the energy density of the background, i.e. the one of the inflaton field,  and the one of the relativistic plasma, when the decay is finished, that is  when ${\Gamma}\sim H_{dec}=H_{kin}\left(\frac{{a}_{kin}}{a_{dec}} \right)^3$,
 will be
\begin{eqnarray}\label{LQIrho}
\rho_{\varphi, dec}=3{\Gamma}^2M_{pl}^2\qquad 
 \mbox{and}  \qquad
\langle \rho_{dec}\rangle=
\langle\rho_{kin}\rangle\left(\frac{{a}_{kin}}{a_{dec}} \right)^3\cong \frac{m_{\chi}^4}{6\pi^2}
e^{-\kappa \alpha m_{\chi}/H}
\frac{\Gamma}{H_{kin}},\end{eqnarray}
\end{enumerate}
one can see that these two constraints bound the decay rate, as follows:
\begin{eqnarray}\label{bound5}
\frac{1}{18\pi^2}e^{-\kappa \alpha m_{\chi}/H_{inf}}\frac{m_{\chi}^4}{H_{kin}M_{pl}^2}\leq \Gamma\leq H_{kin}.
\end{eqnarray}

\

Using, once again,  {\it Stefan-Boltzmann's} law $\rho_{rh}=\frac{\pi^2 g_{rh}}{30}T_{rh}^4$, with  $g_{rh}\cong 106.75$ (the degrees of freedom for the Standard Model), the reheating temperature is given by 
\begin{eqnarray}\label{reheating4}
 T_{reh}=  \left(\frac{30}{\pi^2g_{reh}} \right)^{1/4}
 \langle\rho_{reh}\rangle^{\frac{1}{4}} = 
 \left(\frac{30}{\pi^2g_{reh}} \right)^{1/4}
 \langle\rho_{dec}\rangle^{\frac{1}{4}}
 \sqrt{\frac{\langle\rho_{dec}\rangle}{\rho_{\varphi,dec}}} \nonumber\\
\cong \sqrt{3\pi}\left(\frac{30}{\pi^2g_{reh}} \right)^{1/4}\times e^{-\frac{3}{4}\kappa \alpha m_{\chi}/H_{inf}} \left( \frac{6H_{kin}}{\Gamma}\right)^{1/4}
\frac{m_{\chi}^3}{M_{pl}^2H_{kin}}
M_{pl},
\end{eqnarray}
where we have used that reheating happens when $\langle \rho_{reh}\rangle\sim \rho_{\varphi, reh}$, and recalling that the evotion of both energy densities is given by
\begin{eqnarray}
\langle \rho_{reh}\rangle=\langle \rho_{dec}\rangle
\left(\frac{a_{dec}}{a_{rh}}\right)^4 \qquad
\mbox{and}\qquad \rho_{\varphi, reh}= \rho_{\varphi, dec}
\left(\frac{a_{dec}}{a_{rh}}\right)^6,
\end{eqnarray}
which means that $\left(\frac{a_{dec}}{a_{rh}}\right)^2=
\frac{\langle \rho_{dec}\rangle}{\rho_{\varphi, dec}},$ and thus
\begin{eqnarray}
\langle \rho_{reh}\rangle=\frac{\langle \rho_{dec}\rangle^3}{\rho_{\varphi, dec}^2}.
\end{eqnarray}

\

Finally, taking now into account the bound (\ref{bound5}) we get that for $\alpha\sim 10^{-2}$  the reheating temperature ranges between {   $10^7$ GeV and $10^9$ GeV}. 
It is interesting to
{{compare this enormously high reheating temperature with the solar surface temperature, which  is around $6,000$ K $\sim 8\times 10^{-11}$ GeV (although it is a couple of orders of magnitude higher in its interior).}}

\

\subsubsection{Decay after the end of kination}

In this situation we have the constraint $\Gamma\leq H_{end}$, where $H_{end}$ denotes the end of kination.
Taking this into account, one gets 
\begin{eqnarray}\label{31}
H^2_{end}=\frac{2\rho_{\varphi, end}}{3M_{pl}^2}
\end{eqnarray}
and \begin{eqnarray}  \rho_{\varphi, end}={\rho}_{\varphi,kin}\left( \frac{{a}_{kin}}{a_{end}} \right)^6=
\frac{ \langle{\rho}_{kin}\rangle^2}{{\rho}_{\varphi,kin}},
\end{eqnarray}
where we have used that kination ends when ${ \langle{\rho}_{end}}\rangle\sim{{\rho}_{\varphi, end}}$, meaning that
$\left({a}_{kin}/a_{end} \right)^3=
\frac{\langle{\rho}_{kin}\rangle}{{\rho}_{\varphi,kin}}$, because the energy density of the inflaton decays as $a^{-6}$ but the one of matter as $a^{-3}$. So, the condition ${\Gamma}\leq H_{end}$ leads to the bound 
\begin{eqnarray}\label{bound6}
\Gamma\leq 
\frac{\sqrt{2}}{18\pi^2}e^{-\kappa \alpha m_{\chi}/H_{inf}}\frac{m_{\chi}^4}{H_{kin}M_{pl}^2} .
\end{eqnarray}
Assuming, as is usual, that  thermalization is nearly instantaneous, reheating occurs when the decay is finished, i.e., when $\Gamma\sim H$, and thus $\langle \rho_{dec}\rangle\sim 3\Gamma^2M_{pl}^2$, what  leads to the following reheating temperature
\begin{eqnarray}
T_{reh}=\left( \frac{30}{\pi^2 g_{reh}} \right)^{1/4}\langle\rho_{dec}\rangle^{1/4}= \left( \frac{90}{\pi^2 g_{reh}} \right)^{1/4}\sqrt{{\Gamma}M_{pl}},
\end{eqnarray}
and using the previous bound, one gets that, when de decay of the superheavy particles is after the end of the kination phase,  the reheating temperature belongs in the range 
{ $1 \mbox{  MeV}\leq T_{rh}\leq 10^8$ GeV}.

\section{Historical notes}
 
 \begin{enumerate}
\item  
On Nov. 20, 1915, Hilbert gave a talk at the Royal Society of Science in Göttingen, which was later published in the Transactions of the Society, in March, 1916. There, Hilbert presented the covariant equations for GR. 
On the other hand, Einstein's presentation, for the first time, of his equations for GR took place in the Prussian Academy of Sciences in Berlin on Nov. 25, 1915, this is five days later.

So, at first  glance, it would seem that it was Hilbert the first who obtained these equations. This was indeed the viewpoint of some  scientist contemporary to Einstein and Hilbert, among them Felix Klein, Wolfgang Pauli and Herman Weyl. In fact, for years there was an ongoing controversy about whether it was {Albert Einstein} or {David Hilbert} who first had obtained the GR equations.

\

However a new document appeared later in Hilbert's archive at the University of Göttingen: the printing proofs of the first version of Hilbert's paper published in March of 1916. These proofs were sent to Hilbert two weeks after his talk (on Dec. 6, 1915); and there, one can check that Hilbert did not present the equations of GR in his talk of Nov. 1915. Quite on the contrary, in these proofs Hilbert refers explicitly to Einstein's talk of Nov. 25, published on Dec. 2, 1915. What really happened is that Hilbert included  in his publication the GR equations, which he obtained in an alternative way, but only after reading Einstein's paper, and having made sure to check that the results coincided. 

\

More to the point, Einstein wrote to Hilbert: {\it I had no difficulty finding the general covariance equations of GR.
This is easy with the help of the Riemann tensor. What is really difficult is to recognize that
 these equations constitute a generalization, and even more, a simple and natural generalization of Newton's laws.}

\

What is also clear is that Hilbert discussed GR in a superficial way only, concentrating on the mathematical structure of the equations and on their Lagrangian formulation, but probably without understanding in depth their physical meaning, quite the opposite of Einstein's approach to this issue.

\item
Astronomers who made most important contributions to understand the expansion of our Univers were Vesto Slipher, Henrietta Leavitt and Edwin Hubble. However, the first person who clearly realized that the Universe is expanding, was 
 a Belgian priest, mathematician and physicist, named {Georges  Lema\^\i tre}, who published his results in 1927. We should note that, in those years everybody believed that the universe was static, for very strong physical reasons. Indeed, as any ordinary physical system that had had more than enough time to evolve (infinite, in theory, since the Universe was considered to have always existed), it should have necessarily reached the stationary state. It could not be otherwise. But Lema\^\i tre proved quite the contrary! on the basis of Einstein's GR and matching the theory with the astronomical observations of Slipher and Hubble, in a masterful way, he proved that the Universe was expanding; and later, he observed that it was not eternal, that it had  an origin. 
 
 During many decades (even now, it is still so declared in most places) people believed that it was the astronomer  {Edwin Hubble} the person who first discovered the expansion of the Universe. Only recently, without denying at all Hubble's important contributions, historians have put things in the right place. A very detailed  account of this thrilling story can be found in a book recently published by one of the authors \cite{eliz21}.
 
\item 
In his calculations, using his table of distances (obtained in part with the help of Leavitt's law), and Slipher's table of velocities (obtained as optical Doppler shifts), {Hubble} got a rather large value for the expansion rate, of $H_0\cong 500$  $\frac{{\mbox{km/s}}}{{\mbox{Mpc}}},$ which is off the presently accepted value by almost one order of magnitude. The reason is that it is extremely difficult to measure cosmological distances. On the contrary, obtaining velocities by means of the  Doppler shifts is somehow easier. However, here there is also the problem of appropriately disentangling the recession redshift from other contributions to the observed Doppler shift, coming from the gravitational influence of other massive celestial objects. {Even today, there is  still  a sharp controversy about the right value of $H_0$} (see, e.g., \cite{ekom20} and references therein). Results from astronomical observations by different groups, each one reporting uncertainties of just 1 or 2\% differ by some 5 to 10\%, an unpleasant situation termed the "Hubble parameter tension" \cite{tension1}.

 \item In 1922,  {Alexander Friedmann} was the first to discover full families of solutions of the EEs, which he rightly interpreted as corresponding to expanding and to contracting universes.
 In 1922 and  1924  {Friedman} published two seminal papers in the prestigious German journal  {\it Zeitschrift f\"ur Physik} \cite{friedm12}, of which {Einstein} was an Editor. In those papers, {Friedmann} showed
 that there were solutions to {Einstein's}  equations where the universe evolved in an expanding or contracting  way. Recall (see above) that, in this epoch, everybody believed that our Universe  was static. However, Friedmann explicitly declared in 1924 that, based on some of his solutions, our Universe might well be expanding.
 
 {Einstein} was the 'referee' (this figure did not actually have the same meaning and function at that time) of Friedmann's first paper and, after having studied it, he wrote a letter to the journal saying that  {Friedmann's} calculations should not be published since they contained an error. When { Friedmann} (indirectly) learned of this opinion, he sent all the details of his calculation to Einstein, asking him to check them for himself. After some discussions (described in detail in e.g. \cite{eliz21}) Einstein finally recognized that {Friedmann} did no mistake, and hurried to publish another letter  recognizing his own error and saying that Friedmann's paper should be published. Unfortunately, { in those days it took a long time to see an article published after it was finished}, and { Friedmann} died in 1925 before the publication of {Einsteins'} retraction.

 \item  During a break at the very famous Solvay meeting of 1927 in Brussels,  { Einstein} said literarilly to  {Lema{\^\i}tre} (as was reported later by the last): 
 {''Vos calculs sont corrects, mais votre physique est abominable´´.} He was referring to {Lema{\^\i}tre's} paper of the same year 1927 \cite{lemai27}, he had handed to Einstein during a previous conversation. In the paper (in French) {Lema{\^\i}tre} had obtained for the first time ever the Hubble law and, moreover, he had interpreted it in the right way, as being a proof of the expansion of the Universe, of the very fabric of the cosmos (an interpretation that Hubble never admitted, in his whole life). Einstein himself did not accept the expansion of the cosmos until 1932, when he was finally convinced by Richard Tolmann and Willem de Sitter \cite{eliz21}.
 
 \item 
 In 1956, George Gamow wrote in Scientific American,  that Einstein had told him, long ago, that the idea of the cosmic repulsion associated with the cosmological constant had been "the greatest blunder of his life" (“Die grösste Eselei meines Lebens,” in German). For years to come, this was the only testimony of such a claim; what led many to question it, because of Gamow's well-known imaginative character. Recently, however, it has been discovered by historians of physics (\cite{oraife1} and references therein) that Einstein made a similar statement on at least two more occasions. Indeed, John Wheeler wrote in his book, "Exploring Black Holes: Introduction to General Relativity", that he had personally been present when Einstein said the above words to Gamow, outside the hall of the Institute for Advanced Studies in Princeton. Moreover, Ralph Alpher also testified once that he had heard Einstein make such claim. 
 
 In addition, it is a proven fact that Einstein never wanted to use the CC again, not even when someone suggested that it might be interesting to put it back, so to better adjust the age of the Universe to the results of observations of the oldest galaxies, which seemed at one point (erroneously), to clearly exceed the age of the Universe. As Einstein explained in a footnote in the appendix to the second edition of his book “The meaning of relativity” \cite{94}:
“If Hubble's expansion had been discovered at the time of the creation of the general theory of relativity, the cosmological constant would never had been added. It now seems much less justified to add a term like this in the field equations, since its introduction loses the only justification it originally had.”

Finally, an important consideration, which very few mention, is the following \cite{eliz21}: in taking this position, Einstein was even more radical than Friedmann and Lemaître (the defendants of the expanding universe!), since those always included the CC term in their models for the Universe; even if such term {\it was not necessary} at all in their equations, contrary to the case of Einstein's static model, where it was crucial (for an expanding universe solution, such an additional term plays a secondary role). Anyhow, if there was no reason for its presence, it should not be put there, under any circumstance; this is what Einstein said (please, see \cite{eliz21} for additional details). 

 \item 
{{Fred Hoyle} an English nuclear physicists and astronomer, who formulated the theory of stellar nucleosynthesis (and, with it, the remarkable fact that we are all stardust), was one of the authors of the {\it  Steady State Theory} of the Universe, an attempt to maintain a static model for the cosmos able to account for Hubble's empirical law of expansion \cite{eliz21}.  Hoyle did not buy  Lema{\^\i}tre's conclusion that the universe  had an origin, much less his {\it hypothesis of the primeval atom} that latter exploded; which he, as a serious nuclear physicists, understood it lacked any physical rigor.  On the BBC radio's Third Program broadcast on 28 March 1949, Hoyle explained to the audience that, when comparing Lema{\^\i}tre's model (by then  improved by Gamow) with his steady state theory (where a smooth creation of matter had to take place in order to compensate for Hubble's expansion and keep the matter density of the Universe constant), in Lema{\^\i}tre's model a sudden creation of all the matter in the universe had to occur at the very beginning of it. And, for this to happen, an unbelievably huge expansion (a Big Bang) of the fabric of space was absolutely necessary. But, of course, such phenomenon was fully impossible! and, therefore, he pronounced these famous words in a very disdainful tone (see a much more detailed explanation in \cite{eliz21}).
Hoyle anticipated the idea of cosmic inflation very clearly, albeit as an impossible thought, exactly thirty years before Alan Guth, on an inspired night, could formulate it precisely.

But even if Hoyle had spoken these two words, Big Bang, in a disrespectful manner --trying on purpose to mock Lema{\^\i}tre's model (which had, by then, substantially been improved by George Gamow)-- from this moment on everybody, starting with Gamow himself (a very peculiar character, as is well known) began to use this term to refer to the origin of the Universe.
 
 }
 
 \item 
 The cosmological horizon problem (aka the homogeneity problem) is a fine-tuning issue that affects classical Big Bang models of the universe. It arises due to the impossibility of explaining the homogeneity reported by astronomical surveys of very distant regions of space --which are causally disconnected in these Big Bang models-- unless one invokes a mechanism that sets the same initial conditions everywhere with very high accuracy. This problem  was first pointed out by Wolfgang Rindler in 1956. The most commonly accepted solution is cosmic inflation, as we have discussed here, but an explanation in terms of a variable speed of light has been proposed, too.
 
 \item 
 The flatness problem is another important issue that appeared in the old, classical Big Bang model of the Universe. It was first mentioned by Robert Dicke in 1969, in the Jane Lectures he gave for the American Philosophical Society that year. The total normalized energy density of our present Universe has been measured to be very close to 1, with very small uncertainty, what points towards a very flat universe. Any departure from the conditions leading to this value in the past would had been magnified enormously over cosmic time. This leads to the conclusion that one would need an unbelievably accurate fine tuning in the initial conditions of the Universe, with an energy density that should have been incredibly closer to the critical value at the very beginning of the Universe. As we have seen, cosmic inflation provides the solution to this problem on making the Universe extremely flat, to the needed precision.
 
 \end{enumerate}

\section{Conclusions}

Our main aim in this review was to explain important issues in Modern Cosmology in a  simple and comprehensible way. To start, from Hubble's law and by using an "homogeneous" version of the Einstein-Hilbert action together with the first law of thermodynamics, we have easily derived  the constituent equations of Cosmology. Then we dealt  with their mathematical singularities, namely the famous Big Bang singularity and some possible future singularities, such as the Big Rip and others, providing different ways to remove them and eventually obtain physically sound results. 

\

Next, we have explained about a number of shortcomings  that had appeared in the old Big Bang Cosmology, such as the horizon and the flatness problems, whose  solution was given in terms of the inflationary paradigm introduced by Alan Guth. We have reviewed inflation in detail and in very understandable terms. We have focused on the slow-roll regime, explicitly showing  its attractor behavior and its precise relation with the slow-roll parameters.
We have also calculated, step by step, the number of e-folds the Universe must necessarily expand, in order  to overcome all these problems.

\

Later, we have focused on the study of the current cosmic acceleration via the introduction of different sorts of dark energy. Specifically, we have first considered the most simple model, which uses the cosmological constant. Here we have shown, as a warm up exercise, that the original approach by Einstein --namely his static model of the Universe-- was unstable (in fact, we show in detail that it corresponds to a saddle point of the model, viewed as a dynamical system). Then, we have considered a quintessence field, which can also be used to unify both periods of inflation, under the form of a most popular theory, named quintessential inflation, which we have discussed in some detail.

\

We  have also analysed the Universe's reheating mechanism, which is a very important and  necessary stage after inflation ends; by analysing again a basic system: the quantum harmonic oscillator with a time dependent frequency. Within this simple example, we have  introduced the standard diagonalization method, based on the calculation of the time-dependent Bogoliubov coefficients. As an application, we have derived the bounds of the  reheating temperature for the model of an $\alpha$-attractor in the context of quintessential inflation.

\

To finish, in a closing section we have added a few short notes, which provide updated descriptions of a number of important historical events. A much more detailed account of them is given in \cite{eliz21},  which is a perfect complement to the present, more technical  albeit still very pedagogical review.  

\section*{Acknowledgments}

JdH is supported by grant MTM2017-84214-C2-1-P funded by 
MCIN/AEI/10.13039/501100011033 and by "ERDF A way of making Europe''. EE is supported by MICINN (Spain), project PID2019-104397GB-I00, financed by the State Research Agency, and by the program Unidad de Excelencia María de Maeztu CEX2020-001058-M. Both authors are also supported in part by the Catalan Government, AGAUR project 2017-SGR-247. 
\

\

\end{document}